\documentclass[twocolumn]{aastex631}
\graphicspath{{./}}
\usepackage{amsmath}
\usepackage{appendix}

\usepackage{hyperref}
\usepackage{booktabs}
\usepackage{multirow} 

\begin{document}

\title{Kinematic Mapping of Giant Arcs: A New Method to Locate Lensing Critical Curves}

\author[0009-0009-5611-4924]{Ruwen Zhou}
\affiliation{Department of Physics, 366 Physics North MC 7300, University of California, Berkeley, CA 94720, USA}
\affiliation{School of Physics and Technology, Wuhan University, Wuhan 430072, China}
\affiliation{Tsung-Dao Lee Institute, Shanghai Jiao-Tong University, Shanghai, 520 Shengrong Road, 201210, People's Republic of China}

\author[0000-0003-2091-8946]{Liang Dai}
\affiliation{Center for Astronomy and Astrophysics and Department of Physics, Fudan University, Shanghai 200438, P.R.China}
\affiliation{Department of Physics, 366 Physics North MC 7300, University of California, Berkeley, CA 94720, USA}

\author[0000-0002-3947-7362]{Lingyuan Ji}
\affiliation{Berkeley Center for Cosmological Physics, University of California, Berkeley, CA 94720, USA}
\affiliation{Department of Physics, 366 Physics North MC 7300, University of California, Berkeley, CA 94720, USA}

\author[0000-0002-2282-8795]{Massimo Pascale}
\affiliation{Department of Physics \& Astronomy, University of California, Los Angeles, 430 Portola Plaza, Los Angeles, CA 90095, USA}

\author[0000-0001-9065-3926]{Jose M. Diego}
\affiliation{Instituto de Física de Cantabria (CSIC-UC). Avda. Los Castros s/n. 39005 Santander, Spain}

\author[0000-0002-4622-6617]{Fengwu Sun}
\affiliation{Center for Astrophysics, Harvard \& Smithsonian, 60 Garden St., Cambridge, MA 02138, USA}

\author[0000-0001-7440-8832]{Yoshinobu Fudamoto}
\affiliation{Center for Frontier Science, Chiba University, 1-33 Yayoi-cho, Inage-ku, Chiba 263-8522, Japan}

\correspondingauthor{Liang Dai}
\email{dailiangpku@gmail.com}

\begin{abstract}

The vicinity of lensing critical curves features highly magnified portions of lensed galaxies. Accurate knowledge of the location and shape of the critical curve is useful for understanding the nature of highly magnified sources near critical curves and for revealing dark matter substructures in the lens. In galaxy-cluster lenses, however, prediction of critical curves can be uncertain due to complexity in global mass modeling.
We explore and validate a kinematics-based method for locating the critical curve. This method leverages the continuous line-of-sight velocity profile of the lensed galaxy mapped through integral field spectroscopy of emission lines, and combines an agnostic local lens model and a disk rotation model. Applying our method to a highly magnified region of the Dragon Arc in the Abell 370 cluster lensing field using archival VLT/MUSE IFU mapping of the H$\beta$ line, we constrain the critical curve to an uncertainty band with a half-width of $0.23\arcsec$ ($1\sigma$). This result reveals locations of recently detected extremely magnified stars biased toward the negative-parity side of the critical curve, as predicted for intracluster microlensing. With future JWST/NIRSpec IFU mapping of the H$\alpha$ line at SNR $\simeq 10$ (20), uncertainty could improve to $0.12\arcsec\,(0.08\arcsec)$. A measurement of this type with sufficiently small uncertainty may reveal small-scale wiggles in the shape of the critical curve, which can arise from the lensing perturbation of sub-galactic dark matter substructure. Our approach is generally applicable to caustic-crossing giant arcs and can be incorporated into global lens modeling.

\end{abstract}

\section{Introduction}
\label{sec:intro}

One of the most spectacular situations of gravitational lensing involves a background galaxy overlapping a lensing caustic cast by a foreground galaxy or galaxy-cluster lens. Such a caustic-crossing galaxy often appears as a giant arc, which consists of multiple lensed images joining along lensing critical curves. 

The vicinity of the critical curve is of particular interest in the context of strong lensing studies, as it features a highly magnified portion of the giant arc. The lens map in this vicinity is often approximated by the fold model, which can be quantified using several parameters without having to involve full complexity in the global mass distribution of the lens~\citep{SchneiderEhlersFalco1992textbook, Blandford1986FermatCaustics,Wagner:2019orn}. In this local fold description, a source element close to the caustic on the interior side is mapped into a pair of highly magnified images that straddle the critical curve, with approximate mirror symmetry across the critical curve. The most extremely magnified sources have small sizes and are uncovered near critical curves. For example, $10^2$--$10^3$-fold magnifications have enabled the detection of lensed individual luminous stars~\citep{Miralda1991CausticCrossingStars} in $z\sim 0.7$--$2$ galaxies with the Hubble Space Telescope (HST)~\citep{Kelly2018NatAsM1149, Rodney2018NatAsM0416Transients, Chen2019LensedStarM0416, Kaurov2019LensedStarM0416, Kelly2022FlashlightDozenStars} and the James Webb Space Telescope (JWST)~\citep{Diego2023ElGordoQuyllar, Yan2023PEARLStransients, Fudamoto2025A370DragonLensedStars} imaging. Those are identified as photometric transients due to intracluster microlensing~\citep{Venumadhav2017CausticMicrolensing, Diego2018DMUnderMicroscope, Oguri2018CausticMicrolensing, Diego2019ExtremeMagnification, Weisenbach2024MLnearMacroCausticsReview}. They will allow us to characterize individual massive stars in the young Universe~\citep{Windhorst2018PopIIICausticTransit,Han2024HighlyMagnifiedStar, Lundqvist2024Spectroscopy, Zheng2025BinaryCausticCrossing}, study massive star populations and initial mass functions~\citep{Diego2024AbundanceLensedSupergiantsSpock, Li2025LensedStarsIMFm0416, Li2025arXiv250617565L, Palencia2025Warhol}, and reveal invisible small-scale dark matter structures~\citep[e.g.][]{Dai2018Abell370,Dai2020QCDAxionMinihalos, Williams2023FlashlightsDMSubhalo,Mueller2025PBHlimits, Broadhurst2025FuzzyDM}. Additionally, candidate lensed individual stars with persistent fluxes have been discovered in higher-$z$ caustic crossing galaxies~\citep{Welch2022EarendelNature, Diego2023Mothra, Furtak2024jwstNIRSpecLensedStar, Choe2024Godzilla}, although some may be clusters of stars~\citep[see e.g.][]{Pascale2024Godzilla, Ji25SubhaloEffectOnCaustics, Pascale2025EarendelStarCluster} given their lack of large photometric variability~\citep{Dai2021StarClusterMicrolensing}. Moreover, observations of highly magnified stellar clumps near caustics have offered insight into clustered star formation in extremely dense environments across a range of redshifts~\citep[e.g.][]{Pascale2023SunburstLyC, Vanzella2023SunriseSSC, Adamo2024CosmicGems, Pascale2025EarendelStarCluster}. Studying the positions and flux ratios of their lensed image pairs that straddle the critical curve sensitively probes sub-galactic dark matter (DM) sub-structures~\citep{Dai2020SGASJ1226, Griffiths2021HamiltonObject, Perera2025arXiv251104748P}.

\begin{figure*}[t]
\centering
\includegraphics[width=\textwidth]{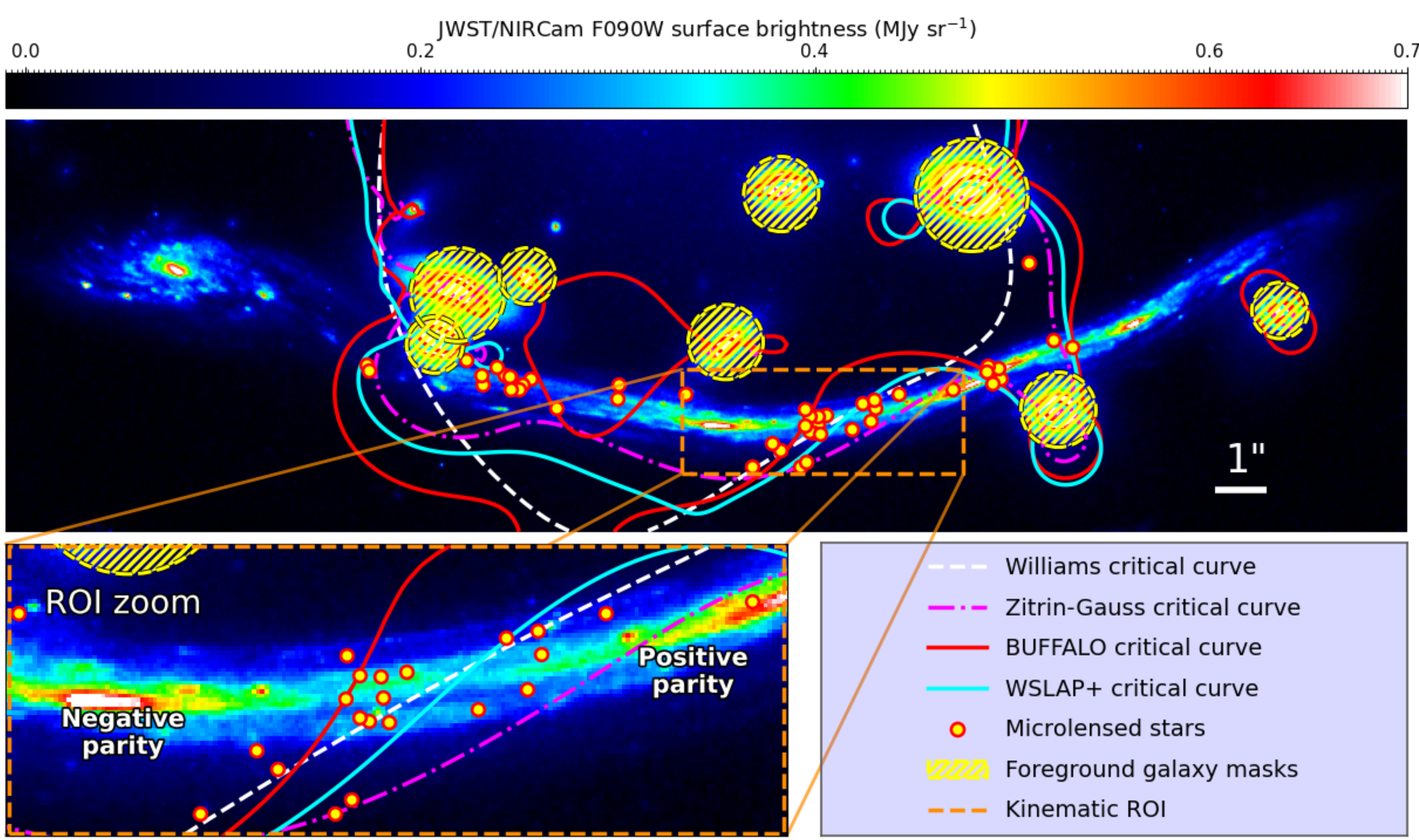}

\caption{JWST/NIRCam F090W surface-brightness map of the Dragon Arc in the lensing field of the galaxy cluster Abell 370. Critical curves at the source redshift $z_s = 0.725$ \citep{Richard2010Abell370} are overlaid for four global lens models: \texttt{Williams} v4.1 \citep[dashed white;][]{William_v4.1_model}, \texttt{Zitrin-Gauss} v1.0 \citep[dash-dotted purple;][]{Zitrin2009lensingltm, Zitrin2013lensingmodel}, \texttt{BUFFALO} v1.0 \citep[red solid;][]{Bufflomodel}, and the hybrid \texttt{WSLAP+} model constrained by JWST data \citep[cyan solid;][]{2025DiegoModel}. The orange dashed rectangle highlights the highly magnified region where the critical curve intersects the giant arc and where two opposite-parity images of the lensed galaxy join together. The lower-left panel zooms into this kinematic region of interest, with negative and positive parity marked along the arc. The lower-right panel summarizes the plotting symbols. Although the surface-brightness morphology provides a qualitative indication of where the critical curve crosses the arc, the precise crossing location remains uncertain.} Yellow filled circles mark a few dozen microlensed red supergiant stars discovered by \citep{Fudamoto2025A370DragonLensedStars}, which are anticipated to appear primarily in the vicinity of the critical curve~\citep{Dai2018Abell370}.

\label{fig:Nircam}
\end{figure*}

The precise location and shape of the lensing critical curve where it intersects a giant arc are of significant value in the study of high magnification lensing phenomena. For microlenses behaving as point masses such as stars or compact objects, theory predicts that micro critical curves fully interconnect within a band on the negative-parity side of the macro critical curve~\citep[Figure 4 of][]{Venumadhav2017CausticMicrolensing}. As a result, microlensed individual stars are more detectable on the negative-parity side than on the positive-parity side. This is also reflected in the probability distribution function of the microlensing magnification factor depending on macro convergence and shear~\citep[e.g.][]{Palencia2024microlensingPDF}. Such behavior of highly magnified stars is seen in the $z=0.94$ caustic-crossing ``Warhol'' arc~\citep{Broadhurst2025FuzzyDM}. If instead an underlying population of diffuse small-scale DM structures dominate microlensing, such asymmetry may be much weaker or absent.

This bias in the spatial distribution of microlensed stars has also been considered for distinguishing between sub-galactic clumpy structures resulting from either wavy dark matter or particle dark matter~\citep{Broadhurst2025FuzzyDM}. Moreover, critical curves are susceptible to perturbations from nearby small-scale DM clumps such as sub-galactic DM subhalos expected in the Cold Dark Matter (CDM) paradigm. It has been shown that across the width of the arc the critical curve may exhibit a wiggly shape owing to substructures~\citep{Dai2018Abell370, Williams2023FlashlightsDMSubhalo, Ji25SubhaloEffectOnCaustics,Broadhurst2025FuzzyDM}.

While the critical curve intersecting the arc can be crudely located based on a mirror symmetry in the surface brightness profile on both sides, a precision determination is far from trivial. Blending of continuum surface brightness features and caustic transients in the lensed galaxy cause ambiguity in identifying lensed image pairs with the smallest separations straddling the critical curve. Magnification perturbations from sub-galactic DM subhalos on small sources may further add confusion to this approach~\citep{Dai2020SGASJ1226, Williams2023FlashlightsDMSubhalo}. Global lens models would ideally predict the exact location of critical curves. In practice, cluster-scale macro models suffer from mass modeling uncertainties, which are especially problematic when it comes to regions of high magnification. To eliminate this problem, model-independent tests have been developed to constrain lens mapping in the vicinity of folds and cusps solely based on comparing multiple lensed images without making any assumptions about the source~\citep{Wagner2017GenModelIndepI, Wagner:2019orn, Wagner:2022ifl}

The Dragon Arc is a spiral galaxy at $z=0.725$~lensed by the galaxy cluster Abell 370, and is one of the first lensed arcs known to astronomers. It has become a prime target for studying extreme magnification phenomena in cluster lensing fields after the recent detection of many highly magnified stars with HST~\citep{Kelly2022FlashlightDozenStars} and JWST~\citep{Fudamoto2025A370DragonLensedStars}. However, independent macro lens models predict critical curve locations that differ at the (sub-)arcsecond level at some places on the arc (see Figure~\ref{fig:Nircam}).

For another example, the locations of critical curve crossings are uncertain between multiple lens models in at least one immensely stretched portion of the $z=2.37$ Sunburst Arc~\citep{Pignataro2021SunburstLensModel, Diego2022Godzilla, Sharon2022SunburstLensModel,SunburstArc_discovery_Rivera-Thorsen17+}, where a highly-magnified young star cluster candidate Godzilla is situated~\citep{VAnzella2020Bowen, Diego2022Godzilla, Pascale2024Godzilla, Choe2024Godzilla}. Similarly, the host galaxies of multiply-imaged Type Ia supernovae in PLCK G165.7+67.0 \citep{Frye2024SNH0pe} and MACS J0138-2155 \citep{Pierel2024Encore} which have been used to measure the Hubble constant \citep{Pascale2025SNH0pe, Suyu2025Encore} lack photometrically identified symmetries that can pinpoint the critical curve to sub-arcsecond precision. The placement of the critical curve affects both time delay and magnification predictions in lens models, which propagate directly to the $H_0$ measurement. Given their importance for cosmology, strong constraints on the critical curve location are valuable anchors to refine these lens systems. All these outstanding challenges motivate the development of empirical methods to reliably locate critical curves to $\sim 0.1\arcsec$ precision or better.

In this work, we exploit kinematic information to measure critical curve locations, and consider the Dragon Arc in Abell 370 as an excellent case study. Our empirical approach complements extensive global lens modeling efforts~\citep{Zitrin2013lensingmodel,William_v4.1_model,Lagattuta2017A370LensModelMUSE,Bufflomodel,Gledhill2024Abell370LensModelCANUCS,Diego2025arXiv250611207D,Eid2025Abell370LensModelFramework}, whose results typically depend on the choice of astrometric constraints as well as lens mass parameterization. With medium resolution ($R\simeq3000$) integral field spectroscopy, the line-of-sight velocity across the lensed arc, which reflects rotation of the galaxy, can be measured from strong nebular emission lines such as H$\alpha$, H$\beta$, [O II]$\lambda\lambda$3726,3728 and [O III]$\lambda\lambda$4959,5007. Compared to stellar photospheric emissions, nebular lines are spatially smoother, and are measurable throughout the entire region of interest (ROI). This source-redshift line selection makes the kinematic observables less susceptible to foreground/background light contamination than broad-band surface brightness morphologies, for which source and cluster emissions are projected onto the same image. Since a static gravitational lens does not change the photon wavelength, the line-of-sight velocity profile exhibits a rough mirror symmetry on both sides of the critical curve along the direction of arc elongation. This general behavior reveals the true location of the critical curve independently of lens mapping far from it. 

While kinematic mapping is a widely practiced technique to study gas dynamics in giant arcs~\citep[e.g.][]{Girard2019AlmaMusekinematics, DiTeodoro2018SNRefsdalhost, Patricio2018KinematicMappingMUSE}, a much smaller body of works have used kinematic information to constrain lens models~\citep{Young2022IFU3DLensReconstruction} and in particular the behavior of critical curves. Focusing on a promising target for the science of highly magnified stars, we apply this method to the Dragon Arc, analyze archival Very Large Telescope (VLT)/Multi-Unit Spectroscopic Explorer (MUSE) data, and assess the potential of using the JWST/NIRSpec integral field unit (IFU).

The remainder of this paper is organized as follows. Section \ref{secdata} summarizes archival data used in this study. In Section \ref{secMethod}, we detail our methodology, starting with an introduction to the principles of velocity mapping of lensed galaxies, followed by descriptions of the disk rotation model and a polynomial local lens model. Section \ref{secvalidation} validates the method via simulations: we construct a simulated Dragon Arc, based on the M74 galaxy, and generate mock observations under both JWST/NIRSpec and VLT/MUSE configurations. Bayesian inference is then applied to constrain the model parameters. Section \ref{sec:results} presents inference on the critical curve from applying the method to both simulated and archival VLT/MUSE data, as well as to simulated JWST/NIRSpec observations. In Section \ref{sec:discussion}, we will discuss the limitations of the current approach, potential improvements, and future applications. Concluding remarks will be given in Section \ref{sec:concl}. In Appendix~\ref{app:a}, we provide a quantitative comparison between our inferred critical-curve band and several independent global lens models. In Appendix \ref{app:b}, we study robustness of our results on the critical curve location against uncertainty in disk rotation modeling.

\section{Data} 
\label{secdata}

This work utilizes archival VLT/MUSE IFU spectroscopic datacubes and JWST/NIRCam imaging of the Abell 370 field, together with a VLT/MUSE IFU spectroscopic datacube of the grand-design spiral galaxy M74 (NGC 0628).

JWST/NIRCam images of Abell 370 were obtained from the CANUCS program (GTO 1208; PI: Chris Willott). Data reduction and mosaic creation were performed by the MAGNIF collaboration using a modified reduction pipeline~\citep{Fudamoto2025A370DragonLensedStars}.

VLT/MUSE IFU spectroscopy data for Abell 370 were retrieved from the ESO archive under programs 094.A-0115(A) (PI: Johann Richard) and 096.A-0710(A) (PI: Franz E. Bauer). These observations were conducted in the WFM-NOAO-N mode, with spatial resolutions limited by seeing. The datacube covers the wavelength range $475$--$953\,$nm at a spectral resolution $R \simeq 3000$ and is sampled at $0.2''$/pixel. The data were reduced with the standard ESO MUSE pipeline (v1.6.1).

We also use VLT/MUSE IFU spectroscopy data of M74 to create mocks. These were obtained from the ESO archive under programs 094.C-0623(A) (PI: Kathryn Kreckel), 095.C-0473(A) (PI: Guillermo A. Blanc), and 098.C-0484(A) (PI: Guillermo A. Blanc). These observations were conducted in the WFM-NOAO-N mode, cover $475$--$935\,$nm at $R \simeq 2830$, and have an average seeing FWHM of $0.66''$ (about $30\,$pc at the distance of M74). The data were reduced with the standard ESO MUSE pipeline (v2.8.1).

\section{Method} 
\label{secMethod}

Our method is best suited to a source galaxy under several conditions: (1) it must straddle a lensing caustic; (2) it should be a star-forming galaxy with spatially extended nebular emission lines; (3) it should possess a rotating disk at a moderate inclination angle.

\subsection{Kinematic Mapping with Nebular Lines}

\begin{figure*}[t]
    \centering
    \includegraphics[width=\textwidth]{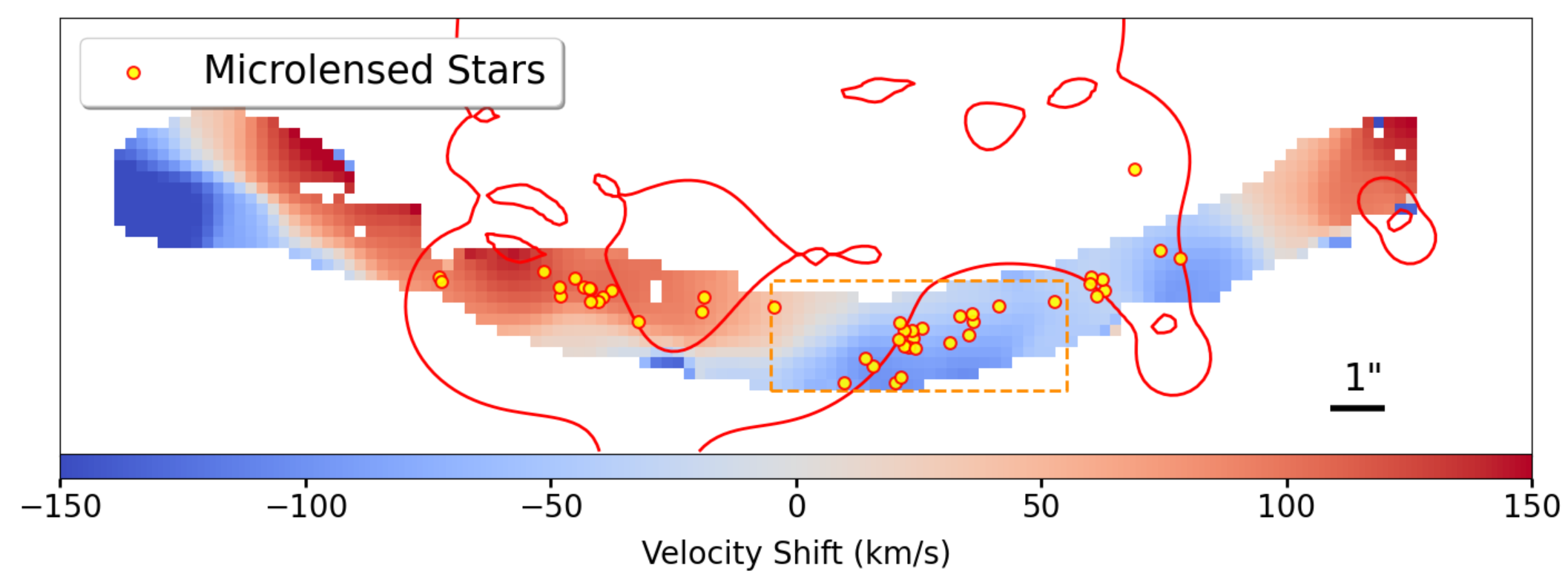}
    \caption{Line-of-sight velocity profile on the Dragon Arc derived from the H$\beta$ line using MUSE IFU data. Red and blue regions trace H{\,\sc ii} regions redshifted and blueshifted with respect to the reference velocity, respectively. Some pixels near the arc edge show anomalous velocity values, likely because foreground light increases the continuum/noise level and degrades the local line fitting near the edge. These pixels lie outside the orange rectangle and therefore do not affect the kinematic inference presented in this work. The red solid line shows the critical curve at $z_s=0.725$ from the \texttt{BUFFALO} lens model~\citep{Bufflomodel}. One region of critical curve intersection, as enclosed by the orange rectangle in Figure~\ref{fig:Nircam}, is where many microlensed individual highly magnified stars are detected through JWST imaging~\citep{Fudamoto2025A370DragonLensedStars}.}
    \label{fig:MUSE_velocity}
\end{figure*}

The disk of the Dragon Arc galaxy is inclined by $45$--$60\,$deg relative to the line of sight. After being mapped onto the image plane, the line-of-sight velocity profile exhibits a clear Doppler-shift pattern dominated by disk rotation. Shown in Figure~\ref{fig:MUSE_velocity}, such a profile is clearly measured from the H$\beta$ line in the archival VLT/MUSE IFU data. The stronger H$\alpha$ line falls outside the wavelength coverage. We adopt H$\beta$ as the primary kinematic tracer instead of the strong metal forbidden doublet [O III]$\lambda\lambda$4959,5007. The H$\beta$ line has a relatively uniform spatial distribution, adequately tracing gas kinematics throughout the entire galaxy. In contrast, the spatial profile of [O III] doublet is more clumpy as a high ionization parameter in the vicinity of hot stars is required to excite doubly ionized oxygen. Looking forward to IFU spectroscopy with JWST/NIRSpec, the redshifted $\mathrm{H}\alpha$ line will be accessible using the G140H/F070LP grating/filter combination.

An additional practical advantage of using source galaxy nebular lines is their separation from foreground cluster light in wavelength space. 
Because the Dragon Arc and the foreground cluster are at different redshifts, the lines used for the velocity measurement appear at the source redshift rather than at the lens redshift; foreground cluster galaxies and intracluster light therefore mainly affect the local continuum level and noise, rather than directly biasing the source emission-line centroid. 
This makes the kinematic observables less susceptible to foreground light contamination than imaging-based approaches relying on broad-band surface brightness morphologies.

\begin{figure}[h]
    \centering
    \includegraphics[width=\columnwidth]{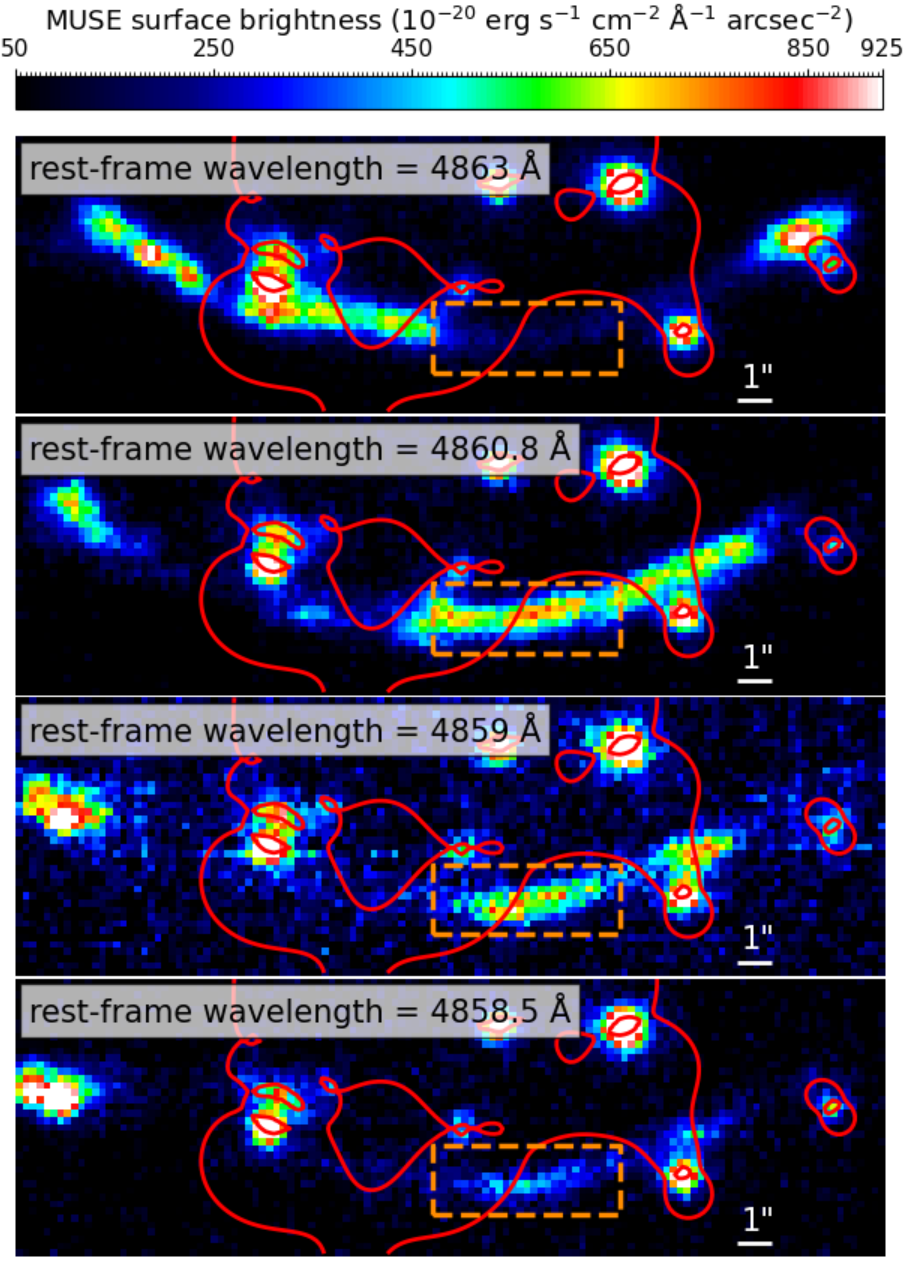}
    \caption{VLT/MUSE IFU data reveals variation of the line-of-sight velocity across the Dragon Arc. Surface brightness varies across the arc as one scans through the $\mathrm{H}\beta$ line from $4863~\AA$ to $4858.5~\AA$ (rest frame). The red solid line shows the critical curve in the \texttt{BUFFALO} lens model~\citep{Bufflomodel}. The behavior of the spatial surface brightness profile reveals roughly symmetric velocity patterns on both sides of the critical curve, as expected from lens mapping near a fold caustic.}
    \label{fig:scan_wavelength}
\end{figure}

Lensing by a static gravitational potential does not change the photon wavelength. Given the typical line widths of H{\,\sc ii} regions $\sigma_v > 1$--$10\,$km/s, the assumption of time independence is a good one. Thus, paired image-plane positions on both sides of the critical curve that map to the same source-plane region should have identical line-of-sight velocities in their emission-line spectra.

In Figure~\ref{fig:scan_wavelength}, we examine the H$\beta$ line within the highly magnified region defined by the orange rectangle in Figure~\ref{fig:Nircam} where the critical curve intersects the arc. When scanning through the H$\beta$ line in wavelength space, bright regions associated with H{\,\sc ii} regions first appear on the east and west ends of the arc and progressively shift toward the critical curve from both directions. Bright emissions eventually disappear on the critical curve as the corresponding H{\,\sc ii} regions hide on the other side of the caustic. This variation in wavelength space reflects the Doppler effect across the disk, while disappearance of brightness on the critical curve is expected from the symmetry of lens map near a fold caustic.

\subsection{Velocity Measurement}
\label{sec:v_measurement}

Mapping of line-of-sight velocity through IFU spectroscopy has been applied to studying gas kinematics in lensed giant arcs~\citep{Girard2019AlmaMusekinematics}. For example, disk rotation of the magnified $z=1.49$ host spiral galaxy of SN Refsdal in the galaxy cluster MACSJ1149.5+2223 is measured based on the [O II] doublet and a flat rotation curve is recovered~\citep{DiTeodoro2018SNRefsdalhost}. Using a similar strategy, \cite{Patricio2018KinematicMappingMUSE} modeled the disk rotation of 8 highly magnified galaxies at $0.6 < z < 1.5$ including the Dragon Arc. However, these previous works did not specifically study lensing critical curves with kinematic mapping, which is the focus of this work.

Prior to measuring the line-of-sight velocity from the IFU datacube, we apply Wiener deconvolution, a classical non-blind deblurring method \citep[see e.g.][]{Wienerdeconvolution, wienerdecon_algorithm}, to obtain an optimal estimate of the intrinsic spectral signal. This preprocessing is applied to each wavelength channel of the datacube. This is essential for mitigating the combined effects of finite telescope aperture and atmospheric turbulence, which cause blurring of the intrinsic line emission profile on the giant arc. Such degradation manifests as reduced spatial resolution and line signal mixing between neighboring source regions. Through mock measurements, we find that skipping preprocessing with Wiener deconvolution causes a bias in the velocity measurement, which leads to a bias in the inferred location of the critical curve.

For VLT/MUSE IFU observations under the WFM-NOAO-N mode, ground-based imaging is significantly compromised by poor seeing.

We approximate the PSF with a Moffat profile~\citep{Trujillo2001MoffatPSF}
\begin{equation}
\text{PSF}(r) = \frac{\beta-1}{\pi\,\alpha^2} \left[ 1 + \left(\frac{r}{\alpha}\right)^2\right]^{-\beta},
\end{equation}
where $\alpha$ is the scale radius of the core, and $\beta$ governs the asymptotic decline of the wings as $\text{PSF}(r) \sim r^{-2\beta}$. We empirically determine  $\alpha \simeq 0.5''$ and $\beta \simeq 1.4$ by measuring unresolved foreground stars in the FoV.

For the JWST/NIRSpec IFU forecasts, we generate a model NIRSpec PSF kernel using \texttt{WebbPSF}~\citep{jwst_Webbpsf}, which allows us to apply Wiener deconvolution with a better-characterized PSF.

With the Wiener-deconvolved datacube, we measure the emission-line center on a pixel-by-pixel basis. For the archival VLT/MUSE data we use H$\beta$, while for the JWST/NIRSpec forecasts we apply the same procedure to H$\alpha$ through the following steps:

\paragraph{1) Continuum subtraction:}
We mask the chosen emission-line core and fit the neighboring continuum to a linear spectrum, which is then subtracted.

\paragraph{2) Peak identification:}
We apply a standard peak-finding algorithm to the continuum-subtracted spectrum using a custom code, selecting the most pronounced line feature near the expected wavelength of the chosen emission line while rejecting noise-dominated peaks.

\paragraph{3) Gaussian fitting:}
We fit the identified line to a Gaussian line profile with a free central wavelength and a free Gaussian width, using the corresponding flux uncertainty estimates as weights in the fitting process.

\paragraph{4) Line-of-sight velocity:}
The best-fit line central wavelength is converted to the line-of-sight velocity relative to the systemic wavelength of the chosen transition using the standard formula for Doppler redshift/blueshift.

\begin{figure}[h]
    \centering
    \includegraphics[width=\columnwidth]{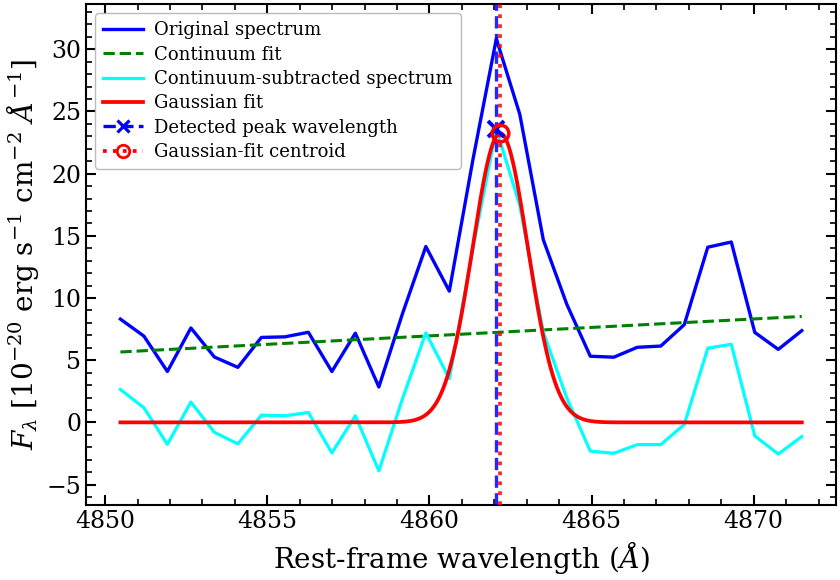}
    \caption{
    Example of Gaussian line-profile fitting to the H$\beta$ emission line in one representative pixel, with continuum subtraction applied. 
    The blue curve shows the original MUSE spectrum, and the dashed green line shows a linear fit to the local continuum. 
    The cyan curve shows the continuum-subtracted spectrum, and the smooth red curve shows the corresponding Gaussian fit. 
    The blue dashed vertical line and blue cross mark the detected peak wavelength of the continuum-subtracted spectrum, while the red dotted vertical line and red open circle mark the Gaussian-fit centroid, which is used to calculate the line-of-sight velocity.
    }
    \label{fig:MUSE_Gaussian}
\end{figure}

\medskip
Figure~\ref{fig:MUSE_Gaussian} shows an example in which we follow these procedures. The approach we outline above outputs a pixelized velocity map (such as the one in Figure~\ref{fig:MUSE_velocity}) robust against spatial blurring and other noise effects.

Figure~\ref{fig:MUSE_velocity} shows the velocity map we derive using the H$\beta$ line in the archival MUSE IFU data. This dataset has a pixel scale of $0.2''$, an effective spatial resolution FWHM$\simeq 0.7''$, and a spectral resolution of R $\simeq 3000$. The median signal-to-noise ratio (SNR) per pixel in the H$\beta$ line is $\simeq 7$.

Low spatial resolution eliminates information about velocity variation on small scales, preventing accurate localization of the critical curve below the smoothing scale and obscuring any possible small-scale curvilinear structure. The archival MUSE IFU dataset used here suffers significantly in this regard.

Spectral resolution is another crucial factor that is especially important when it comes to highly magnified regions adjacent to critical curves. Such regions map to a very small area on the source plane next to the caustic, across which only a small amount of differential rotation is present. That said, even at an instrumental spectral resolution that only modestly resolves the emission line profile, it is still possible to centroid the line center with better precision if higher SNRs per spectral element are available, provided that the line profile model used is robust.

\subsection{Velocity Profile on Image Plane}

We require a parameterized model to fit the measured line-of-sight velocity field on the image plane, particularly in the vicinity of the critical curve. Through parameter inference, this will inform us about the behavior of the critical curve and provide uncertainty estimates. Our model consists of two independent major components: a disk rotation model, and a parameterized model of local lens map independent of global lens modeling.

\subsubsection{Disk Rotation Model}

Given that the Dragon Arc is a spiral galaxy, we consider a disk rotation model to define its primary kinematic structure. This simple model has 5 parameters: ${\rm RA}_0$, ${\rm DEC}_0$, $i$, $\phi$, and $v_{\rm rot}$. The position $({\rm RA}_0,\,{\rm DEC}_0)$ defines the kinematic center of the rotating disk on the source plane. The inclination angle $i$, ranging from $0^\circ$ (face-on) to $90^\circ$ (edge-on), affects the projection of the physical rotation velocity vector along the line of sight. The position angle $\phi$ specifies the intrinsic orientation of the disk apparent major axis in the plane of the sky, measured from North ($0^\circ$) to East ($+90^\circ$). Lastly, $v_{\text{rot}}$ is the circular velocity of disk rotation, which we approximate as independent of the galactocentric distance.

Projecting the circular disk rotation onto the line-of-sight direction, we find the line-of-sight velocity at any point on the source plane
\begin{equation}
v_{\text{los}} = v_{\text{rot}}\,\sin i\, \cos(\theta - \phi).
\end{equation}
Here $\theta$ is the azimuthal position in the disk plane about the galaxy center, which can be easily found for the line of sight passing any point on the source plane and intersecting the disk. We have neglected a random velocity component for the H{\,\sc ii} regions.

In principle, the disk rotation curve can be calculated from a 3D mass model of the spiral galaxy and its host DM halo~\citep{Patricio2018KinematicMappingMUSE}. We assume a constant circular velocity $v_{\rm rot}$ rather than a general radial function. A fairly ``flat'' rotation curve is found in typical spiral galaxies at $3$--$10\,$kpc~\citep{Sofue2001GalaxyRotationCurveARAAreview, Blok2008THINGSGalaxyRotationCurves, Lelli2016DiskGalaxyMassModels}. It decently approximates some strongly lensed spiral galaxy~\citep{DiTeodoro2018SNRefsdalhost}. Moreover, the highly magnified region near the critical curve maps to a relatively small area on the source galaxy, which spans up to a few kpc around the turnover regime of the rotation curve. Over a limited range of galactocentric distance, the rotation curve is approximately flat. This simple choice reduces the number of model parameters, but generalization to a more sophisticated parameterization of the rotation curve, if necessary, will be straightforward.

\subsubsection{Local Lens Model}
\label{sec:lensmodel}

\begin{figure}[h]
    \centering
    \includegraphics[width=\columnwidth]{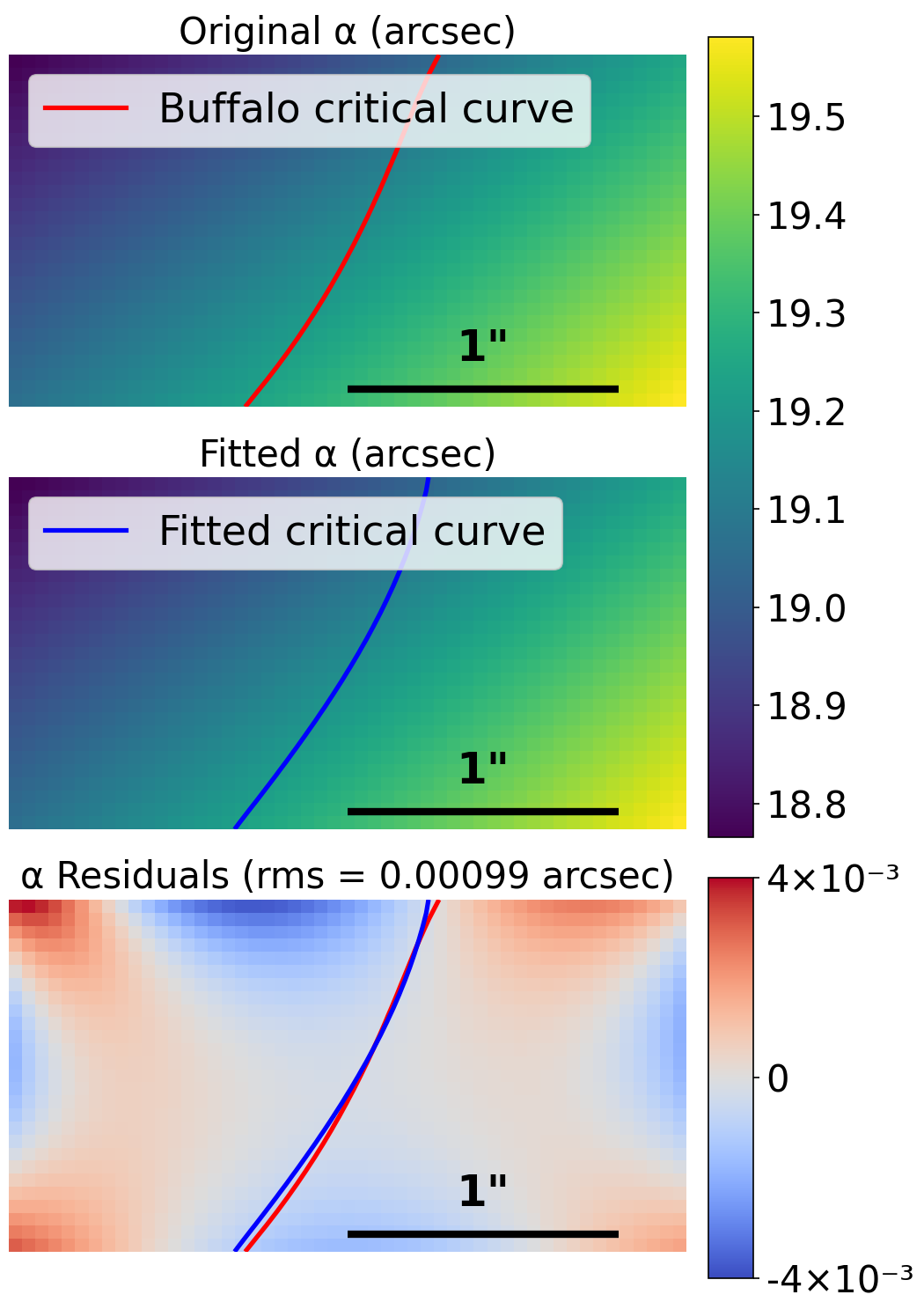}
    \caption{Fitting a 3rd-order polynomial deflection field to the \texttt{BUFFALO} lens model~\citep{Bufflomodel} in a $2.5''\times1.3''$ vicinity of the critical curve. The magnitude of the deflection angle, $\alpha =\sqrt{\alpha_1^{2}+\alpha_2^{2}}$, is color-coded. The upper panel shows the original \texttt{BUFFALO} deflection field and the model critical curve (red), while the middle panel shows the best-fit polynomial deflection field and the corresponding critical curve (blue). The bottom panel shows that the fitting residuals have a root-mean-square (RMS) of $\sim 0.001''$.
    }
    \label{fig:polyfit}
\end{figure}

To sufficiently describe lens mapping between the image plane and the source plane near the critical curves, we employ a third-order polynomial expansion for the deflection field. This approach goes beyond the simple fold caustic model \citep[see e.g.][]{SchneiderEhlersFalco1992textbook, Blandford1986FermatCaustics, Miralda1991CausticCrossingStars}, which has a limited range of validity near the critical curve. To exploit a sufficient amount of kinematic information, we need to include a sizable portion of the arc extending on both sides of the critical curve. On the image plane, significant departure from the simple fold model is therefore expected. A continuous deflection field in the form of a third-order polynomial provides a flexible enough parameterization to capture the general behavior of the local lens map, regardless of the global mass model. 

We work in a local Cartesian coordinate system $(x_1,\,x_2)$ on the image plane, with coordinates measured in angular units. 
Angular positions are measured relative to a reference point $(x_{0,1},\,x_{0,2})$ through $\Delta x_1 = x_1 - x_{0,1}$ and $\Delta x_2 = x_2 - x_{0,2}$. The deflection angle $(\alpha_1,\,\alpha_2)$ is parameterized as
\begin{align}
    \alpha_1 = &\ a_{00} + a_{10}\,\Delta x_1 + a_{01}\,\Delta x_2\nonumber\\
&+ a_{20}\,\Delta x^2_1 + a_{11}\,\Delta x_1\,\Delta x_2 + a_{02}\,\Delta x^2_2\nonumber \\
& + a_{30}\,\Delta x^3_1 + a_{21}\,\Delta x^2_1\,\Delta x_2 + a_{12}\,\Delta x_1\,\Delta x^2_2 + a_{03}\,\Delta x^3_2, \\
    \alpha_2 = &\ b_{00} + b_{10}\,\Delta x_1 + b_{01}\,\Delta x_2\nonumber\\
&+ b_{20}\,\Delta x^2_1 + b_{11}\,\Delta x_1\,\Delta x_2 + b_{02}\,\Delta x^2_2\nonumber \\
& + b_{30}\,\Delta x^3_1 + b_{21}\,\Delta x^2_1\,\Delta x_2 + b_{12}\,\Delta x_1\,\Delta x^2_2 + b_{03}\,\Delta x^3_2.
\end{align}
For a Newtonian gravitational potential, and assuming a single lens plane, the deflection field is the gradient of the lensing potential~\citep{SchneiderEhlersFalco1992textbook}, so the polynomial coefficients are subject to constraints
\begin{align}\label{eq:degenerate}
b_{10} &= a_{01},\nonumber \\
b_{20} &= a_{11}/2,\nonumber \\
b_{11} &= 2\,a_{02},\nonumber \\
b_{30} &= a_{21}/3,\nonumber \\
b_{21} &= a_{12},\nonumber \\
b_{12} &= 3\,a_{03}.
\end{align}

Since $a_{00}$ and $b_{00}$ are degenerate with ${\rm RA}_0$ and ${\rm DEC}_0$, respectively, the original set of 20 polynomial coefficients reduces to 12 independent parameters: $\{a_{10}, a_{01}, a_{20}, a_{11}, a_{02}, a_{30}, a_{21}, a_{12}, a_{03}\}$ for the Cartesian component $\alpha_1$, and $\{b_{01}, b_{02}, b_{03}\}$ for the other Cartesian component $\alpha_2$.

The disk rotation model describes the intrinsic velocity field on the source plane, while the polynomial model for the deflection field determines mapping of the intrinsic velocity profile onto the image plane. These two ingredients are linked through the ray equation~\citep{SchneiderEhlersFalco1992textbook},
\begin{equation}
\label{lens_eq}
    y_i = x_i - \alpha_i(\vec{x}), \quad i = 1,2.
\end{equation}
where $(y_1,\,y_2)$ are angular coordinates on the source plane. Critical curves are found by requiring a degenerate Jacobian matrix $A_{ij}$
\begin{equation}
\det A = \det\left(\delta_{ij} - \frac{\partial \alpha_i}{\partial x_j}\right) = 0.
\end{equation} 

We additionally convolve the model image-plane velocity profile with a two-dimensional Gaussian kernel whose width is a free parameter $\sigma$. While instrumental PSF smearing is treated with Wiener deconvolution, this does not perfectly undo it in the IFU data. We introduce in our model this {\it ad hoc} smearing effect to enhance effectiveness of model fitting.

Figure~\ref{fig:polyfit} shows the effectiveness of a 3rd-order polynomial model when it is employed to approximate the deflection field of the \texttt{BUFFALO} model~\citep{Bufflomodel} in a $2.5''\times1.3''$ region centered on the critical curve.

In particular, we do not intend to make the simplifying assumption that the critical curve is locally a straight line or a circular arc~\citep{Dai2018Abell370}. Sizable curvature or more complex deformations can often be induced by nearby galaxy-scale perturbers~\citep{Dai2020SGASJ1226, Perera2025arXiv251104748P}. Such departures from a simple straight-line critical curve are important to account for in highly stretched tangential arcs such as the Dragon Arc and the Sunburst Arc~\citep{Pignataro2021SunburstLensModel, Diego2022Godzilla}.

One more physical constraint we will implement is derived from the observed morphology of the arc. As can be seen in Figure~\ref{fig:Nircam}, in the orange rectangular region arc elongation has an orientation in the range $[0^\circ,\,15^\circ]$ relative to the horizontal direction (West). This corresponds to the degenerate direction of the local lens mapping, which is set by the eigenvector of the Jacobian matrix $A_{ij}$. This is not enforced in the general polynomial model of ray deflection. We choose to constrain within this conservative range the degenerate direction evaluated at the center of the orange rectangular region, which is in the middle of the arc and very close to the recovered critical curve.

\section{Validation of Method}
\label{secvalidation}

To assess the fidelity of our method, we create mock IFU data that resembles the lensing situation of the Dragon Arc and perform kinematic mapping on the mock data. To that end, we require a nearby spiral galaxy whose nebular emission profile is measured at a spatial resolution well exceeding the level needed for analyzing strongly lensed galaxies at cosmological distances.

\begin{figure}[t]
    \centering
    \includegraphics[width=\columnwidth]{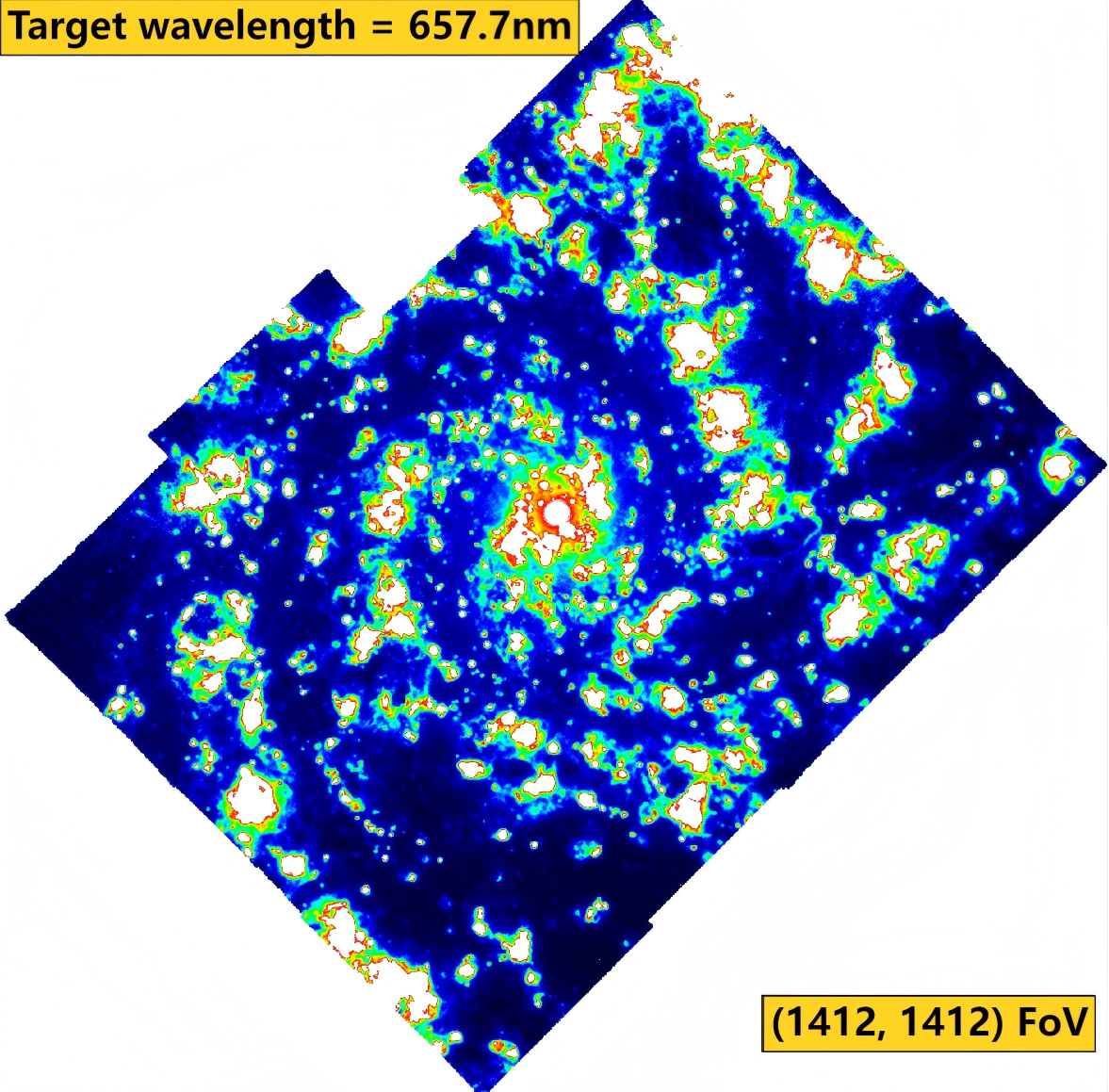}
    \caption{H$\alpha$ (observed at 657.7 nm) image of M74 (NGC 628) resolved down to $\sim 30\,$pc, derived from archival MUSE IFU data. This galaxy is chosen as the source galaxy for creating mocks due to its grand-design spiral morphology, well-resolved H{\,\sc ii} regions, and nearly face-on orientation that simplifies kinematic analysis. Powered primarily by sites of star formation, the underlying spatial profile of H$\alpha$ surface brightness is clumpy. High-quality archival IFU data make it a favorable analog for the Dragon Arc, allowing for detailed forward modeling of lensing effects and instrumental effects.}
    \label{fig:M74_H_beta}
\end{figure}

\subsection{M74}

For the purpose of creating mocks, we choose the local spiral galaxy M74, for which archival MUSE IFU data is available. This choice has several advantages: 

\begin{figure*}[t]
    \centering
    \includegraphics[width=\textwidth]{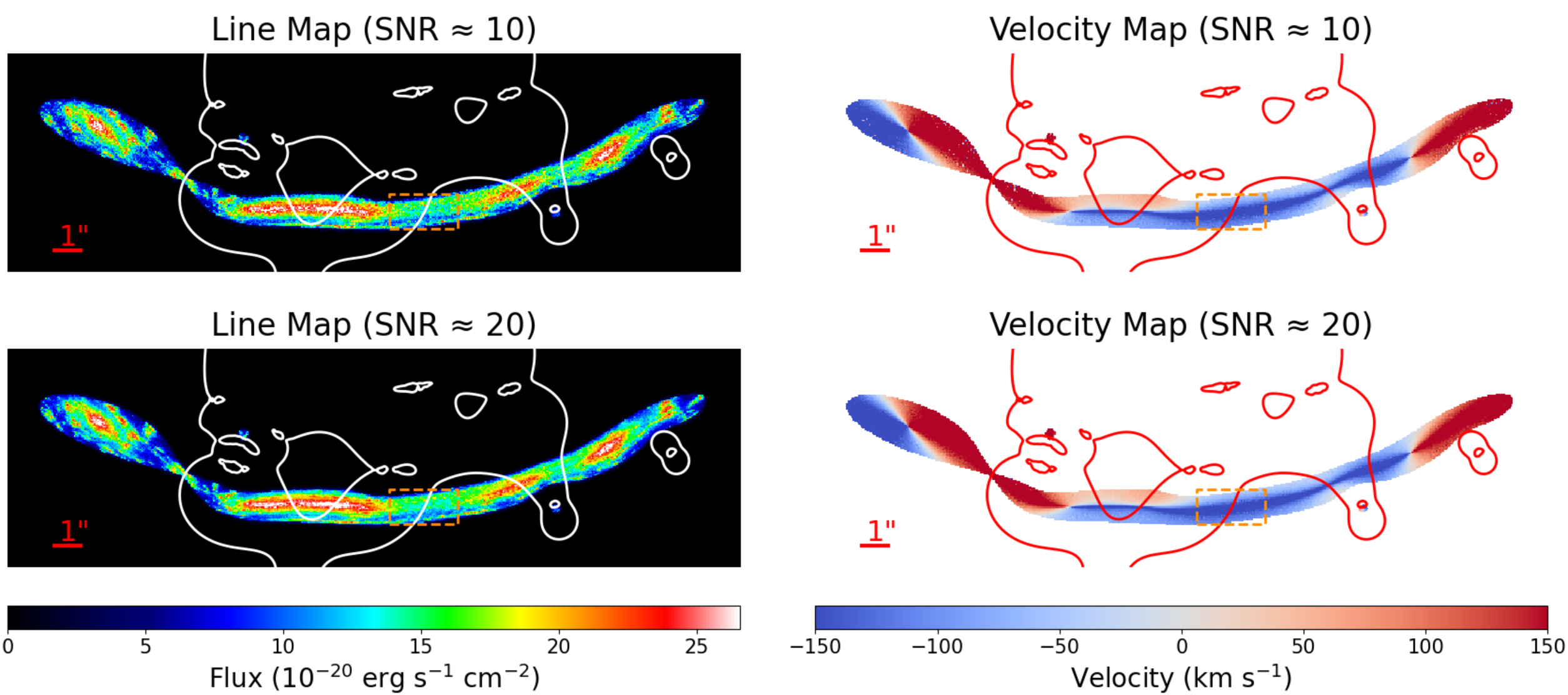}
    \caption{Mock IFU data of the Dragon Arc created for observational conditions appropriate for JWST/NIRSpec. See Section~\ref{secvalidation} for details. This mock data is used to validate the method of kinematic mapping in this work. Panels show flux map at the H$\alpha$ line (left) and the measured line-of-sight velocity (right), at two different noise levels (SNR $\simeq$ 10 and 20). The fold symmetry in the velocity value is visible on both sides of the critical curve. The critical curve (white in the left panels and red in the right panels) is derived from the interpolated \texttt{BUFFALO} model. The orange dashed rectangle (2.5\arcsec $\times$ 1.3\arcsec) indicates the ROI of JWST configuration used for the Bayesian inference.}
    \label{fig:jwst_config}
\end{figure*}

    \paragraph{1) Morphological similarity to Dragon Arc:}
    M74 is a ``grand-design'' spiral galaxy with well-defined spiral arms, making it an excellent analog for the Dragon Arc galaxy.

    \paragraph{2) Well-resolved H{\,\sc ii} regions:}
    The distribution and properties of M74’s discrete H{\,\sc ii} regions (Figure~\ref{fig:M74_H_beta}) help us better understand to what extent nebular emission can be approximated as spatially continuous when a similar galaxy at $z \sim 1$ is imaged at a limited spatial resolution. 

    \paragraph{3) Face-On Orientation:}
    M74 is nearly face-on with a negligible Doppler effect due to disk rotation. Starting with a nearly zero inclination disk, we will artificially implement disk inclination in the synthetic data.

\medskip
We use the public PHANGS-MUSE data of M74 (DR2 public release; catalog ID NGC0628\_PHANGS\_DATACUBE\_native.fits)~\citep{2022A&A...659A.191E}, acquired through 45 pointings over 23 observational blocks (2014-2017). The datacube covers the wavelength range $4759$--$9350\,\AA$ with $\Delta\lambda = 1.25\,\AA$ spectral sampling (R $\simeq 2830$ at the wavelength of H$\alpha$), achieving a median SNR of 30 per spectral pixel in H{\,\sc ii} regions. 

\subsection{Creating an Analog Galaxy}

To create an analog of the lensed Dragon Arc galaxy based on observed nebular emissions in the local spiral M74, we go through the following procedures:

\paragraph{1) Emission Line Extraction:}

We extract the H$\alpha$ line by isolating a narrow wavelength range around it pixel by pixel in the MUSE IFU datacube. Through the method described in Section~\ref{sec:v_measurement}, we perform pixel-by-pixel Gaussian line profile fitting to obtain a spatially well-resolved H$\alpha$ flux map with clearly identified H{\,\sc ii} regions.

\paragraph{2) Implementing Inclination:}

 While M74 is nearly face-on, we artificially implement a disk inclination of $i = 60^{\circ}$. The corresponding Doppler shift in each pixel will be computed assuming a constant rotation curve of $v_{\rm rot} = 200\,$km/s.

\paragraph{3) Mock IFU Spectra:}

For each spatial pixel, we fit a Gaussian H$\alpha$ line to the original M74 spectrum and apply the required Doppler shift. The line width and flux are set accordingly to the original MUSE data, while the line center is shifted according to the disk rotation model and disk inclination. This procedure yields pixel-by-pixel synthetic spectra that consistently encode both disk rotation and intrinsic line properties.

\medskip

The source-plane IFU data produced this way incorporate a realistic spatial distribution of H{\,\sc ii} regions and the prescribed kinematic properties.

\subsection{Mock IFU Data}

\begin{figure*}[htb]
    \centering
    \includegraphics[width=\textwidth]{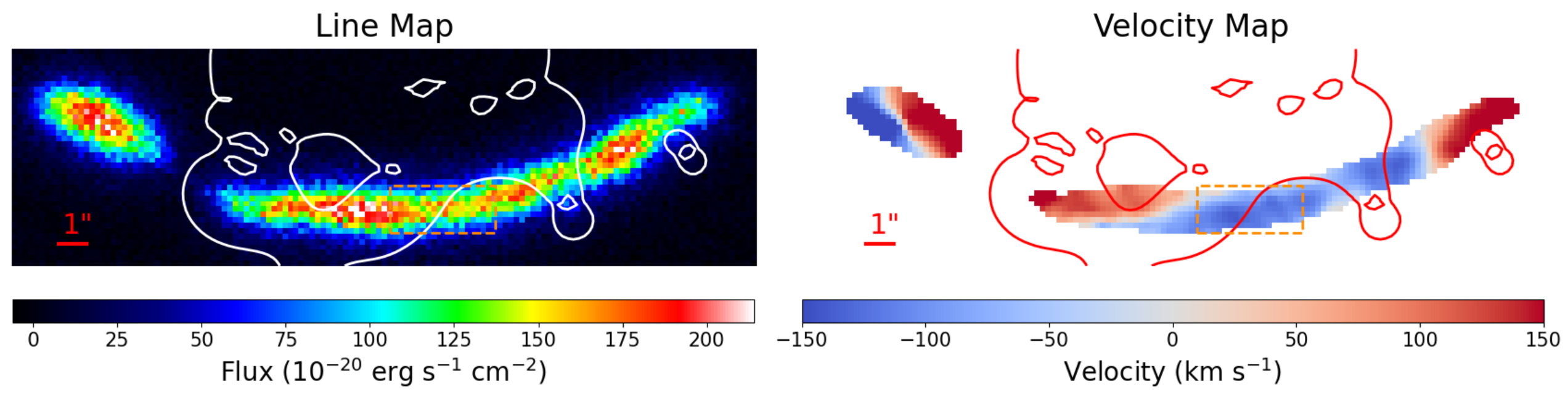}
    \caption{Simulated MUSE IFU data for a lensed spiral that resembles the Dragon Arc. Panels show the H$\beta$ flux per pixel (left) and the measured line-of-sight velocity field (right). The critical curve from the \texttt{BUFFALO} lens model is drawn. Bayesian modeling of disk rotation and lensing will be applied within the orange dashed rectangle (3.8\arcsec $\times$ 1.6\arcsec).}
    \label{fig:muse_config}
\end{figure*}

We need to create a mock giant arc that resembles the Dragon Arc observed with JWST/NIRSpec IFU. We create image-plane data by artificially lensing the mock spiral galaxy on the source plane with the \texttt{BUFFALO} macro lens model~\citep{Bufflomodel}. We adjust the position and orientation of the mock spiral to roughly reproduce the morphology of the Dragon Arc. A NIRSpec PSF model created with the \texttt{Webbpsf} package \citep{jwst_Webbpsf} is then convolved with the image-plane data. To emulate realistic observational conditions, we renormalize the mean line flux to match what is observed in the MUSE data for Dragon Arc (converting H$\beta$ flux to H$\alpha$ flux assuming Case B recombination). We then include synthetic photon shot noise and readout noise to realize two different benchmark noise levels per pixel --- SNR $\simeq 10$ and SNR $\simeq 20$ integrated over the H$\alpha$ line profile. For JWST/NIRSpec IFU, the finalized mock IFU datacube is mapped to a pixel scale of $0.05\arcsec$ (this pixel sampling, finer than the NIRSpec IFU instrumental sampling at $0.1\arcsec$ per pixel, is achievable through a dither pattern) and has a spectral resolution of $R \simeq 3000$. 

Examples of this mock data are shown in Figure~\ref{fig:jwst_config}. The left two panels present the line maps at the two noise levels, while the right two panels display the corresponding velocity maps. These measurements are obtained following the methodology outlined in Section~\ref{sec:v_measurement}. We again emphasize the importance of performing Wiener deconvolution prior to measuring velocity. Using the PSF kernel generated by the \texttt{Webbpsf} tool \citep{jwst_Webbpsf}, deconvolution significantly reduces biases in the final Bayesian inference results. We will further make this point in the following sections.

Despite uncertainty in source galaxy properties and the macro lens model, which renders it difficult to perfectly replicate the Dragon Arc, our simulated arc reproduces many important features, such as topology and morphology of multiple images and fold symmetry. The mock simulations are adjusted to reproduce the photometric and kinematic scales of the Dragon Arc that are relevant for our analysis. 
The mean emission-line surface brightness over the arc region agrees with the MUSE measurement to within $0.25\%$, although the detailed pixel-level distribution is not expected to match because the mock uses the H{\,\sc ii}-region distribution of M74 as a template. 
Within the ROI, the line-of-sight velocity span is $\sim160\,$km/s in the MUSE data and $\sim160$--$180\,$km/s in the mock simulations. 
The mock therefore reproduces the local velocity scale to the $\sim10\%$ level; the remaining difference is mainly an approximately constant velocity offset, which is irrelevant for our kinematic-mapping method because the constraint is driven by local velocity gradients and fold symmetry rather than the absolute systemic velocity.
As expected, the JWST/NIRSpec IFU mock (Figure~\ref{fig:jwst_config}) has a superior quality than the archival VLT/MUSE data (Figure~\ref{fig:MUSE_velocity}), with an improved pixel sampling ($0.05''$ versus $0.2''$) and higher signal-to-noise ratios (SNR $\simeq$ 20 compared to SNR $\simeq$ 7 in MUSE).

For the mock VLT/MUSE IFU observation, we follow a procedure analogous to that in the JWST/NIRSpec case but adapt to the corresponding instrumental characteristics. The MUSE mock (Figure~\ref{fig:muse_config}) has a coarser pixel sampling of $0.2\arcsec$. The PSF is modeled as a Moffat profile (Section~\ref{sec:v_measurement}). Due to the MUSE wavelength coverage in this configuration, we use the H$\beta$ emission line instead of H$\alpha$, scaling the line fluxes appropriately. We assume a moderately lower integrated SNR $\simeq 7$ per pixel integrated over the H$\beta$ line. The mock datacube has a comparable spectral resolution ($R \simeq 3000$) but Wiener deconvolution is less effective due to the coarser sampling and broader PSF.

\subsection{Bayesian Inference}

With mock arcs in hand under both the VLT/MUSE and JWST/NIRSpec IFU configurations, we validate our image-plane modeling of the velocity profile through Bayesian inference. The \texttt{BUFFALO} lens model is adopted as the ground truth for the lensing deflection field, which we aim to recover.

According to Bayes' theorem, the posterior probability distribution of the model parameters $\mathbf{\Theta}$ given the observed data $\mathbf{d}$ is:
\begin{equation}
\mathcal{P}(\mathbf{\Theta} | \mathbf{d}) \propto \mathcal{L}(\mathbf{d} | \mathbf{\Theta})\,\pi (\mathbf{\Theta})
\end{equation}
where $\mathcal{L}(\mathbf{d} | \mathbf{\Theta})$ is the likelihood of observing $\mathbf{d}$ given a model with parameters $\mathbf{\Theta}$, and $\pi(\mathbf{\Theta})$ is the prior probability distribution for $\mathbf{\Theta}$. The likelihood function is defined as:
\begin{equation}
\mathcal{L}(\mathbf{d} | \mathbf{\Theta}) \propto \prod_{i \in \mathcal{M}} \exp\left\{ -\frac{1}{2} \left( \frac{v_{\text{obs},i} - v_{\text{pred},i}(\mathbf{\Theta})}{\sigma_{\text{obs},i}} \right)^2 \right\}
\end{equation}
where $\mathcal{M}$ denotes the set of valid pixelized data points within the diamond-shaped mask region. Here, $v_{\text{obs},i}$ and $v_{\text{pred},i}(\mathbf{\Theta})$ are the observed velocity and the model-predicted velocity, respectively, and $\sigma_{\text{obs},i}$ is the corresponding uncertainty in velocity measurement. 

We have the following set of model parameters:
\begin{equation}
\mathbf{\Theta} = \left\{
\begin{aligned}
& {\rm RA_0}, {\rm DEC_0}, i, \phi, v_{\rm rot}; \\
& a_{10}, a_{01}, a_{20}, a_{11}, a_{02}, a_{30}, a_{21}, a_{12}, a_{03}; \\
& b_{01}, b_{02}, b_{03};\\
& \sigma
\end{aligned}
\right\}\nonumber 
\end{equation}

We employ uniform priors for all parameters with boundaries set sufficiently wide to encompass physically plausible values. However, we impose two important physical constraints on the lens model, as already discussed in Section~\ref{sec:lensmodel}. To ensure efficient and robust exploration of this high-dimensional parameter space, we adopt a two-stage inference strategy:

\paragraph{1) Global parameter optimization:}
We first optimize $\mathbf{\Theta}$ using the differential evolution algorithm (implemented in \texttt{SciPy}). This step yields an initial estimate of the best-fit parameters.

\paragraph{2) Bayesian sampling:}
We then use parameter values found by global optimization to initiate Markov Chain Monte Carlo (MCMC) runs using the \texttt{emcee} package ~\citep{emcee}.

\medskip

We utilize the posterior samples of $\mathbf{\Theta}$ to derive the posterior distribution of the lensing critical curve. Recovery of the ground-truth critical curve from the \texttt{BUFFALO} model within the inference uncertainty will validate our methodology.

\begin{figure}[h]
    \centering
    \includegraphics[width=\columnwidth]{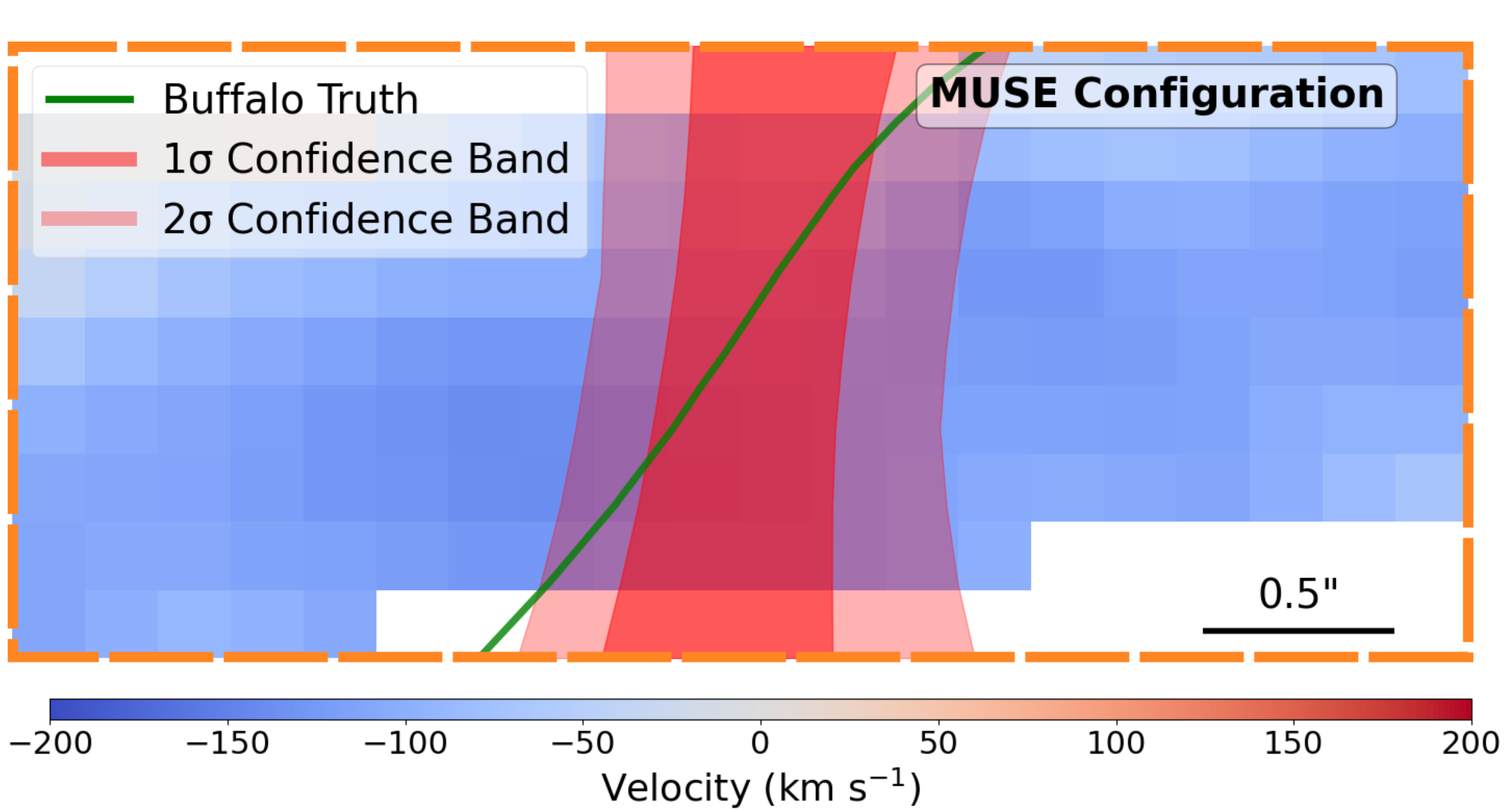}
    \caption{Mock measurements performed for the simulated lensed spiral galaxy and for the observational quality of VLT/MUSE IFU. The mean half width of the 1$\sigma$ (2$\sigma$) confidence band is 0.25\arcsec (0.50\arcsec). While recovery is less accurate near the upper and lower edges due to insufficient pixels, the band is consistent with the true critical curve in the central part of the arc. Compared with the JWST/NIRSpec IFU case, the MUSE IFU results are more uncertain due to the significantly worse spatial resolution.}
    \label{fig:muse_ccband}
\end{figure}

\section{Results}
\label{sec:results}
We will discuss results separately under the two observational conditions considered: VLT/MUSE IFU and JWST/NIRSpec IFU.

\subsection{VLT/MUSE IFU}

Figure~\ref{fig:muse_ccband} presents a simulated measurement on the mock MUSE IFU data, applied to the MUSE-mock ROI indicated by the orange dashed rectangle in Figure~\ref{fig:muse_config}.

Despite the poor spatial resolution, we are able to locate the critical curve to within a confidence band that has a half width of $0.25\arcsec$ at the $1\sigma$ confidence level. The band has an orientation that differs moderately from that of the true critical curve, but is consistent within the posterior distribution.

We note that the quality of constraint from kinematic mapping alone sensitively depends on the specific realization of disk rotation on the source plane near the caustic. In Figure~\ref{fig:muse_ccband}, the velocity variation is relatively modest, and at many places the velocity profile along the direction of arc elongation is rather flat. If the line-of-sight velocity varied more strongly across that region, the local lens map and the critical curve would be better constrained. Additional tests on JWST mock data indicate that a 25\% increase in the characteristic velocity variation can improve the inferred critical-curve constraint by about 12.5\%, while a 50\% increase can improve it by about 25\%. The exact scaling relations are not universal, but depend also on SNRs, spatial resolution, and the specific lensing configuration.

\begin{figure*}[htb]
    \centering
    \includegraphics[width=\textwidth]{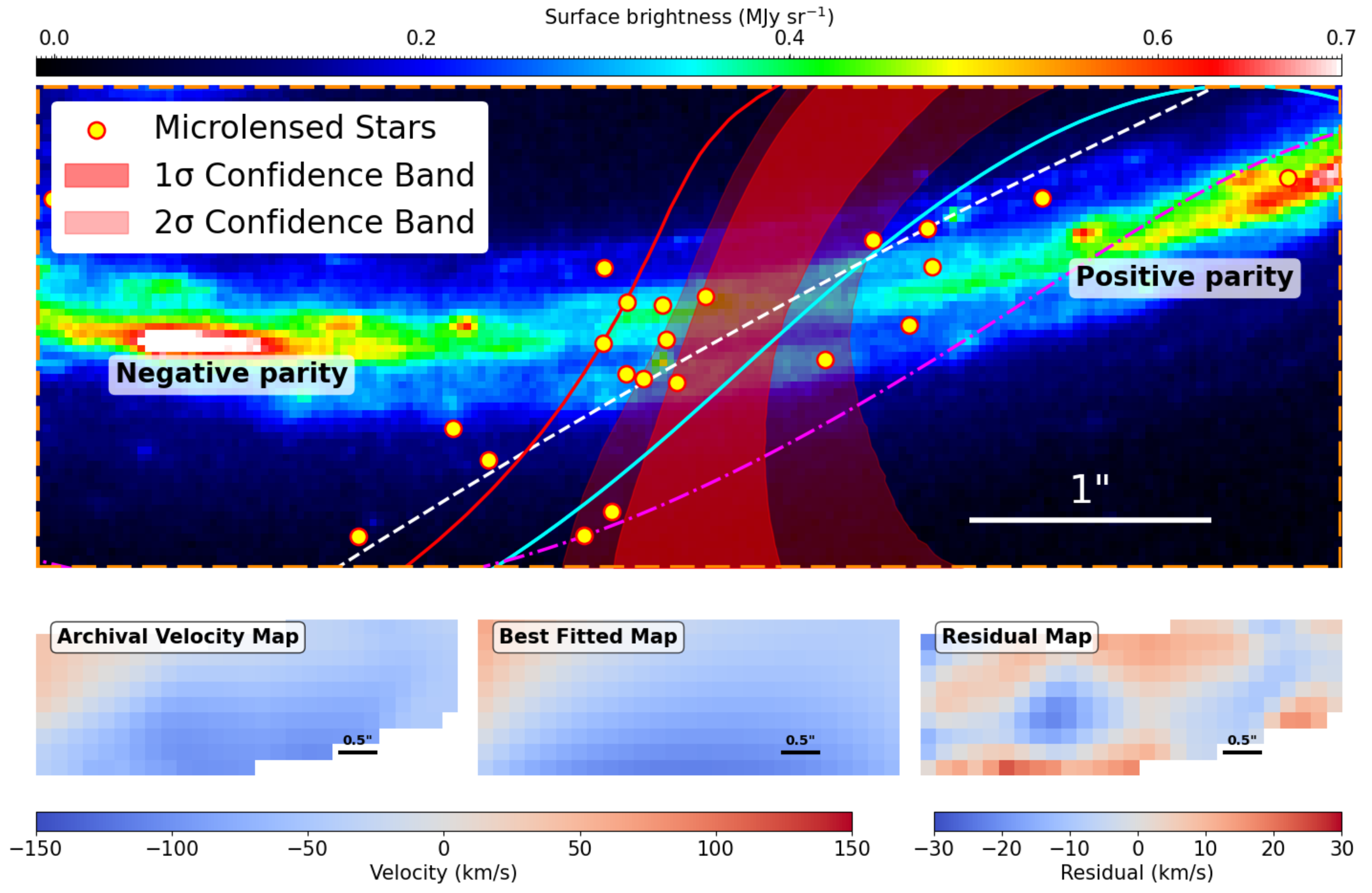}
    \caption{
    \textbf{Upper panel:} JWST NIRCam F090W image zoomed into the orange rectangular region marked in Figure~\ref{fig:Nircam}, overlaid with the confidence bands for the critical curve derived using the H$\beta$ line from the archival MUSE data. The mean half width of the 1$\sigma$ (2$\sigma$) confidence band is 0.23\arcsec (0.48\arcsec). Critical curves predicted by the same four global lens models as described in Figure~\ref{fig:Nircam} are also drawn: \texttt{Williams} v4.1 \citep[dashed white;][]{William_v4.1_model}, \texttt{Zitrin-Gauss} v1.0 \citep[dash-dotted purple;][]{Zitrin2009lensingltm, Zitrin2013lensingmodel}, \texttt{BUFFALO} v1.0 \citep[red solid;][]{Bufflomodel}, and the \texttt{WSLAP+} model \citep[cyan solid;][]{2025DiegoModel}.
    \textbf{Lower panel:} \textbf{(left)} Image-plane line-of-sight velocity profile measured from the archival MUSE IFU data, \textbf{(middle)} the best-fit image-plane velocity profile, and \textbf{(right)} the corresponding residuals. }
    \label{fig:archival}
\end{figure*}

In Figure~\ref{fig:archival}, results are derived by applying our methodology to the real MUSE IFU data. We constrain the critical curve with a posterior half width of $0.23\arcsec$ at the $1\sigma$ confidence level. If we relax the constraint on the local degenerate direction, the $1\sigma$ half width inflates to $0.30\arcsec$. As expected, the critical curve is more tightly constrained in the middle of the arc than in the upper and lower outskirts.

The residuals shown in Figure~\ref{fig:archival} are at the level of $\sim 10\,$km/s and are spatially correlated on $\sim0.5$--$1\arcsec$ scales. This may indicate a deficiency in the assumed disk rotation model, or distortions in the local lens map not captured by the polynomial model (which may be induced by sub-galactic DM subhalos), or unknown data analysis systematics. However, information is inadequate to draw robust conclusions.

Before comparing our result with independent global lens models, we emphasize that our kinematic-mapping method is designed primarily to constrain the location and shape of the critical curve, rather than to uniquely determine the local mass distribution. 
The polynomial lens model used in our inference is intentionally flexible without imposing any specific global mass parameterization. 
It therefore does not by itself break lens-map degeneracies such as the mass-sheet degeneracy~\citep{MSD_Schneider2013}. 
Under a mass-sheet transformation with scaling parameter \(s\), the convergence \(\kappa\), shear \(\gamma\), and lensing Jacobian \(A\) transform as
\[
\kappa' = s\kappa + (1-s), \qquad
\gamma' = s\gamma, \qquad
A' = sA ,
\]
It follows that \(\det A'=s^2\det A\). 
Thus the critical curve, defined by \(\det A=0\), is invariant under the transformation, while the local convergence, shear normalization, and absolute mass normalization are not. The kinematic-mapping method therefore sidesteps this degeneracy by targeting the critical-curve location itself. That said, such degeneracies will be broken by physical priors if kinematic lens constraints are incorporated into global mass modeling.

Our derived confidence band has an orientation most similar to that of the critical curve predicted by the \texttt{BUFFALO} lens model~\citep{Bufflomodel}, but is displaced significantly westward relative to it. 
Interestingly, our inferred critical-curve location appears more consistent with the symmetric pattern of the continuum surface brightness, although this information is not used in our inference.

Critical curves predicted by the two other models shown in Figure~\ref{fig:Nircam} make shallower angles with respect to the arc, and one of them is less compatible with the fold symmetry of the surface-brightness profile. 
As quantified in Appendix~\ref{app:a}, our confidence band is most consistent with the non-parametric \texttt{WSLAP+} lens model~\citep{Broadhurst2025FuzzyDM, Diego2024DMSmallestScales}, which is augmented by a compilation of multi-image systems with spectroscopic redshifts~\citep{Gledhill2024Abell370LensModelCANUCS} and is also constrained by a dozen multiply-imaged surface-brightness features identified across the Dragon Arc in deep HST and JWST images~\citep{Broadhurst2025FuzzyDM}.

Taking the middle of the confidence band to be our derived critical curve, about two-thirds of the microlensed highly magnified stars discovered in \cite{Fudamoto2025A370DragonLensedStars} fall on the side of negative parity. If the \texttt{BUFFALO} model critical curve is adopted, most microlensed stars would fall on the side of positive parity. The bias to negative parity is qualitatively consistent with microlensing by intracluster stars~\citep{Venumadhav2017CausticMicrolensing}, but we do not intend to compare to quantitative models as statistics are low here. Detection of many more microlensed stars in the Dragon Arc from upcoming JWST/NIRCam multi-epoch imaging (GO 7345; PI: Y.~Fudamoto) will mitigate Poisson fluctuations to allow quantitative studies.

From the posterior distribution of polynomial lens models, we compute the local gradient of the eigenvalue of the Jacobian matrix, defined as $\mathbf{d} = - \nabla(\kappa + \lambda) = (|\mathbf{d}|\sin\alpha,\ -|\mathbf{d}|\cos\alpha)$~\citep{Venumadhav2017CausticMicrolensing}. On the critical curve, $\mathbf{d}$ has a magnitude $|\mathbf{d}| = 7.7 \pm 1.3~\text{arcmin}^{-1}$ and $|\mathbf{d}|\sin\alpha = 7.4\pm 1.6~\text{arcmin}^{-1} $, reflecting the angle $\alpha$ between the critical curve and the degenerate direction.

The local convergence inferred from our posterior spans \(\kappa \simeq 0.75\)--\(0.79\). For comparison, the local convergence in the same region from the macro lens models shown in Figure~\ref{fig:Nircam} spans \(0.61\)--\(0.67\) for the \texttt{BUFFALO} model, \(0.70\)--\(0.76\) for the \texttt{Zitrin-Gauss} model, \(0.60\)--\(0.66\) for the \texttt{Williams} model, and \(0.61\)--\(0.65\) for the \texttt{WSLAP+} model. The offset between our local posterior and the global models should be interpreted in light of the lens-map degeneracies discussed above, while the spread among the global models reflects differences in their parameterizations, input constraints, and reconstruction methods.

\begin{table}[htbp]
\centering
\caption{Derived disk parameters for the Dragon Arc: inclination $i$, position angle $\phi$, and rotation velocity $v_{\rm rot}$. We measure the position angle from North ($0^\circ$) to East ($+90^\circ$). We compare to \cite{Patricio2018KinematicMappingMUSE}, which is based on the global lens model of \cite{Lagattuta2017A370LensModelMUSE}.}
\label{tab:parameter_results}
\begin{tabular}{cccc}
\toprule
Parameter & Median & 68\% C.I. & \cite{Patricio2018KinematicMappingMUSE} \\
\midrule

$i$ [$^\circ$] & $70$ & $(57,\,78)$ & $54\pm 1$ \\
$\phi$ [$^\circ$] & $-38$ & $(-43,\,-32)$ & $-30\pm 3$ \\
$v_{\rm rot}$ [km~s$^{-1}$] & $304$ & $(267,\,329)$ & $V_{\rm max} = 207\pm 7$ \\
\bottomrule
\end{tabular}
\end{table}

Our derived $v_{\rm rot}=270$--$330\,$km/s is larger than the typical range for spiral galaxies $200$–$250\,$km/s. While we find a disk orientation largely consistent with \cite{Patricio2018KinematicMappingMUSE}, our inferred $v_{\rm rot}$ is significantly larger than the maximal velocity $V_{\rm max}$ there (Table~\ref{tab:parameter_results}). This discrepancy is attributed to several factors. First, we have adopted a simple flat rotation curve rather than one based on a physical galaxy mass model. More importantly, while in \cite{Patricio2018KinematicMappingMUSE} disk parameters are recovered from the northeast lensed image of the full disk (see Figure~\ref{fig:Nircam}), our local analysis in the orange rectangle fits only a part of it, which does not cover locations of extremal line-of-sight velocity values. In fact, we have set a broad prior $150$–$350\,$km/s for $v_{\rm rot}$ to thoroughly explore the space of the local polynomial deflection field, but the posterior distribution piles up near the high end of the $v_{\rm rot}$ prior, which reflects the limitation of kinematic fitting restricted to a part of the disk. 

As we show in Appendix~\ref{app:b}, this limitation in recovering $v_{\rm rot}$ does not compromise the ability to locate the critical curve. When we restrict $v_{\rm rot}$ to a narrow prior $195$–$205\,$km/s, a similar posterior distribution of critical curves is derived, consistent with results derived without the tight prior on $v_{\rm rot}$. For the purpose of constraining the critical curve, the source-plane velocity field model does not have to be ``physical''. As long as the model can well fit the velocity field in the relevant source-plane region, fold symmetry guarantees additional robust information to constrain local lens mapping.

\begin{figure}[h]
    \centering
    \includegraphics[width=\columnwidth]{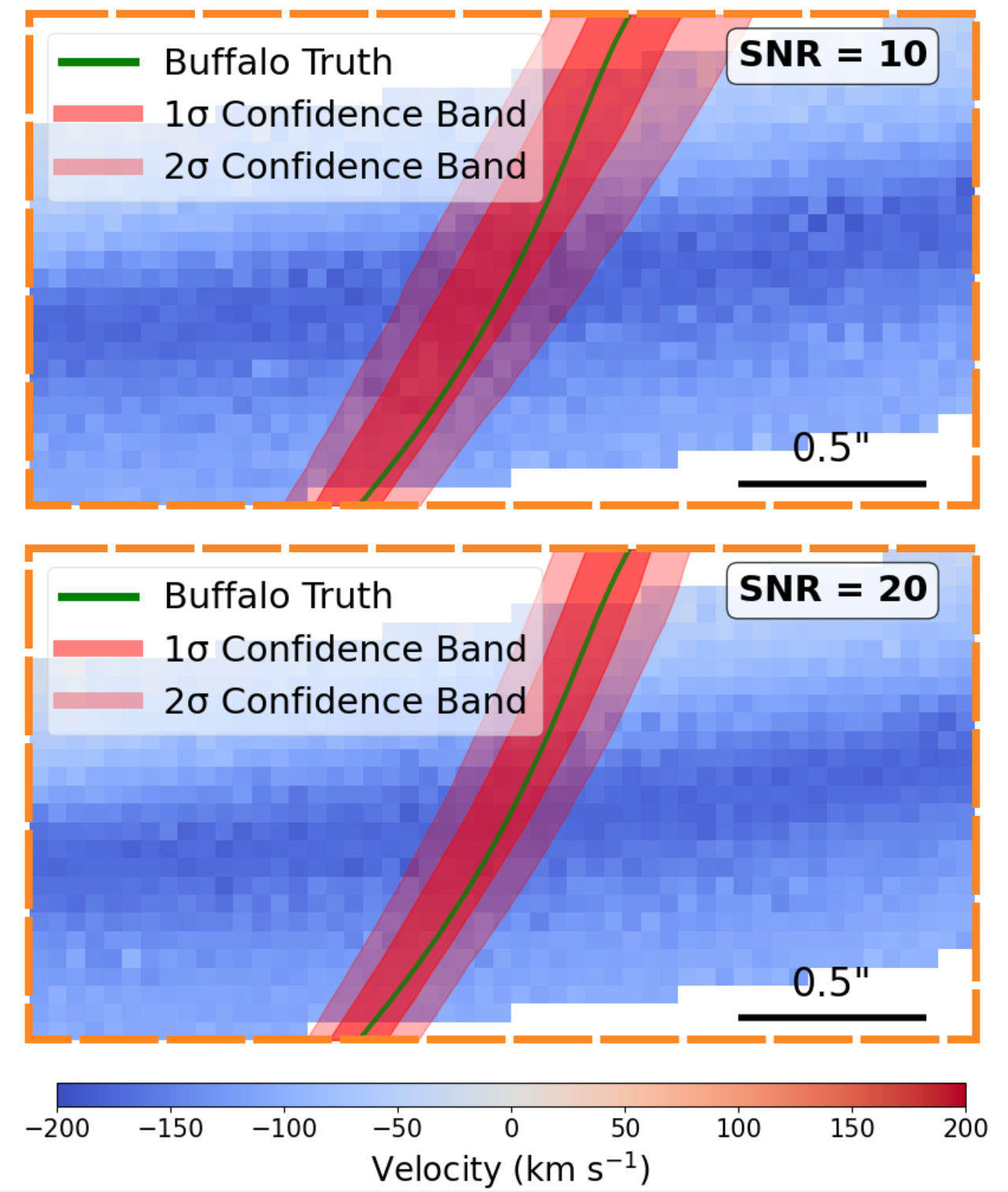}
    \caption{Constraints on the critical curve location from mock JWST/NIRSpec IFU measurements, derived for the same simulated lensed spiral galaxy that resembles the Dragon Arc as in Figure~\ref{fig:muse_ccband}, and at two different H$\alpha$ mean SNR levels (per pixel). Color-coded pixels show the line-of-sight velocity field (km/s) measured for the orange rectangular region shown in Figure~\ref{fig:jwst_config}, with red bands representing the 1$\sigma$ (dark red) and 2$\sigma$ (light red) confidence intervals. The green solid line shows the ground truth, which is the critical curve from the \texttt{BUFFALO} lens model. For SNR=10, the mean half width of the 1$\sigma$ (2$\sigma$) confidence band is 0.12\arcsec (0.22\arcsec). For SNR=20, this improves to 0.08\arcsec (0.16\arcsec). In this example, the true position of the critical curve is recovered without bias.}
    \label{fig:jwst_ccband}
\end{figure}

\begin{figure}[htb]
    \centering
    \includegraphics[width=\columnwidth]{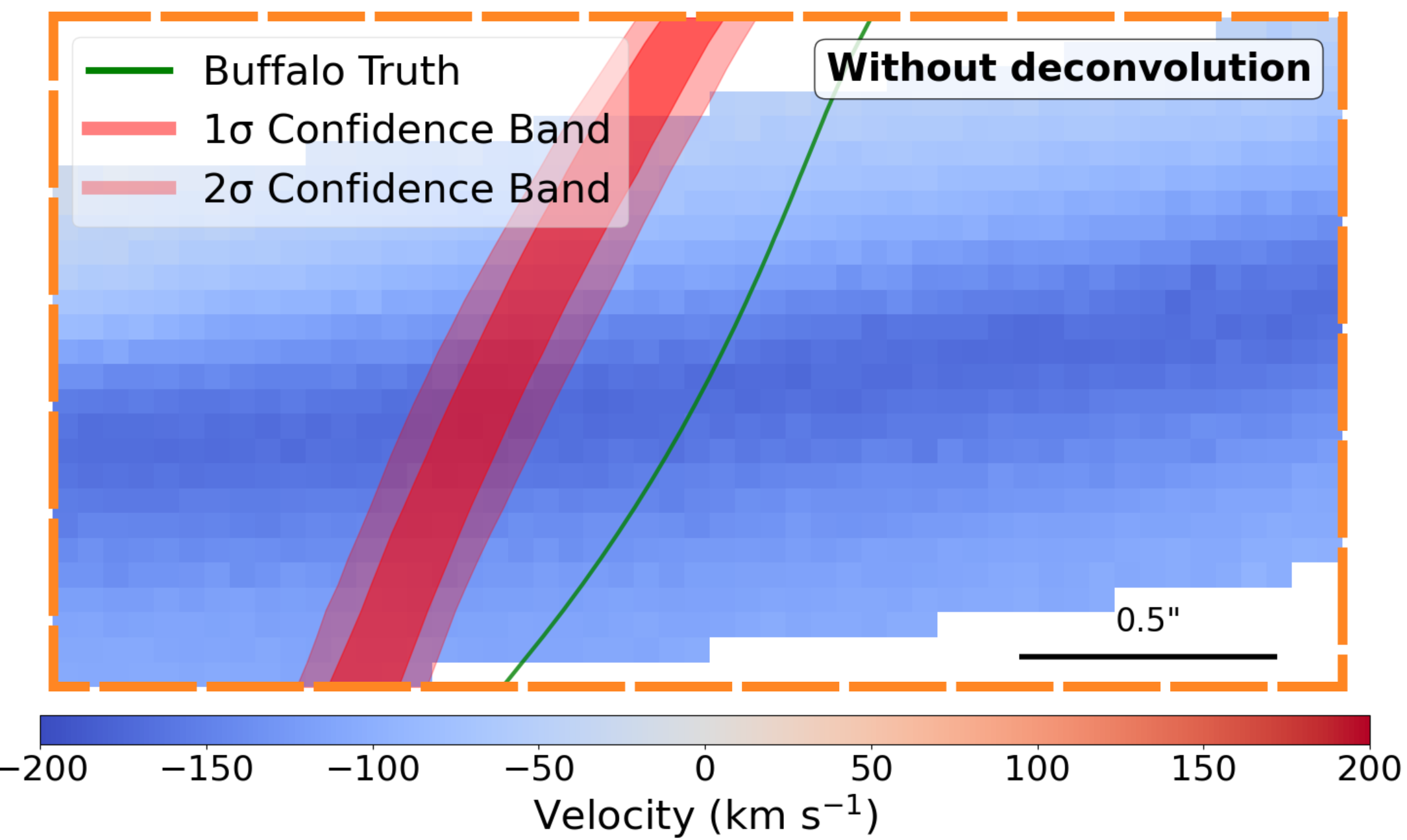}
    \caption{Mock JWST/NIRSpec IFU measurement corresponding to the SNR $\simeq$ 20 case in Figure~\ref{fig:jwst_ccband}, but with velocities measured from IFU images without PSF deconvolution. In contrast to the results shown in Figure~\ref{fig:jwst_ccband}, Bayesian inference locates the critical curve with a large bias. In our method, PSF deconvolution proves essential in eliminating bias at VLT/MUSE and JWST/NIRSpec IFU spatial resolutions.}
    \label{fig:nodeconvolution}
\end{figure}

\subsection{Forecast for JWST/NIRSpec IFU}

Figure~\ref{fig:polyfit} shows that a polynomial model fits the \texttt{BUFFALO} deflection field with negligible residuals $<0.005\arcsec$ within the $2.5\arcsec \times 1.3\arcsec$ vicinity of the critical curve. A major difficulty, however, arises from the intrinsically shallow velocity profile across this region. The subtle velocity variation is easily washed out at the poor spatial resolution of MUSE, with the region having only $12.5 \times 6.5$ pixels. To mitigate this, we have to fit the MUSE-derived velocity field in a larger field of view $5.4\arcsec \times 2.0\arcsec$ (Figure~\ref{fig:archival}). Although the enlarged fitting region more effectively constrains the critical curve location, the effectiveness of the polynomial deflection model degrades, resulting in relatively large velocity residuals.

We expect significant improvement if the same region is mapped with JWST/NIRSpec IFU, for which the PSF is smaller by a factor of $\sim 6$. For the same intrinsically shallow velocity profile, a smaller PSF and a finer pixel-sampling will allow the same gradient to be measured more accurately, and the polynomial deflection model may be used in a smaller vicinity of the critical curve.

Mock measurements performed under such conditions are presented in Figure~\ref{fig:jwst_ccband}, where the confidence band for the critical curve has a half width of only $0.12\arcsec$ ($0.08\arcsec$) at the $1\sigma$ level if the H$\alpha$ line is measured at SNR $\simeq 10$ ($20$), a factor of $\simeq 2$--$3$ improvement upon MUSE IFU. According to the JWST Exposure Time Calculator~\citep[ETC;][]{JWSTetc2016}, about 8 hr (29 hr) of exposure is required to achieve SNR $\simeq 10$ ($20$) per pixel.

The limitation will be dominated by the finite spectral resolution as well as systematics in modeling the source-plane velocity profile and the local lens mapping.

With the superb imaging quality of JWST/NIRSpec IFU, Wiener deconvolution or similar deconvolution algorithms will be a crucial step of preprocessing prior to reliable velocity measurements. Figure~\ref{fig:nodeconvolution} shows that without deconvolution our method yields a tight constraint on the critical curve location, which is however significantly biased from the ground truth.

\section{Discussion}
\label{sec:discussion}
In this section, we discuss limitations, potential improvements, and possible extensions of this work.

\paragraph{1) Additional Local Lens Constraints:}
We have so far applied two constraints on the local lens map. The first one, Eq.~\ref{eq:degenerate}, is the requirement of the lensing potential. The second one is an observational constraint on the elongation direction of the arc. In addition to these, image pairs of highly magnified stars identified in the future can be incorporated as additional constraints, although confidently identifying the image pair of the same lensed star may be non-trivial.

\paragraph{2) Improved Disk Rotation Model:}
Our disk rotation model assumes a constant circular velocity $v_{\rm rot}$. While this is a simplification, it is well-motivated for the outer disk regions probed by the highly magnified arc, given the flat rotation curves observed in typical spiral galaxies at radii beyond a few kpc~\citep{Sofue2001GalaxyRotationCurveARAAreview, Blok2008THINGSGalaxyRotationCurves, Lelli2016DiskGalaxyMassModels}. Nevertheless, one can readily incorporate more complex and realistic rotation curves, such as one derived for an exponential disk embedded in a Navarro-Frenk-White (NFW)~\citep{CDMhalo1996} dark matter halo. Additionally, the velocity model may allow a random velocity component for the H {\sc ii} regions reflecting ISM turbulence.

\paragraph{3) Improved Velocity Measurements:}
We have used a simple linear model for local continuum fitting and assumed a Gaussian profile for emission lines. Both treatments can be improved for more physically realistic measurements. First, the stellar continuum could be modeled more rigorously using dedicated tools such as the PPXF code together with a library of stellar spectra, following the approach of \cite{Patricio2018KinematicMappingMUSE}. This may be particularly relevant in the case of the hydrogen Balmer lines for which a narrow nebular emission line is superimposed on top of a broad photospheric absorption line. 

These effects could in principle influence the local continuum estimate, the detailed line profile, and hence the inferred peak location in individual spectra, and should therefore be regarded as caveats of the present implementation. However, they represent limitations of the specific line-measurement procedure adopted here rather than of the kinematic-mapping methodology itself. More sophisticated treatments of the stellar continuum and line profile will therefore make velocity measurements more robust. Secondly, the assumption of a Gaussian line profile may be relaxed.
\paragraph{4) Generalization:}
The applicability of our method is fundamentally rooted in its use of kinematic information, making it viable for any giant arc mapped with IFU spectroscopy. This allows us to utilize a variety of spectroscopic tracers across different wavebands: optical emission lines from H {\sc ii} regions for star-forming galaxies, photospheric absorption lines for quiescent galaxies, or molecular lines (e.g. with ALMA) for dusty, gas-rich galaxies. The spectral information is not limited to the line center, but may also include parameters describing the line shape. Additionally, it will be powerful to incorporate kinematic information in global lens modeling and combine spectral information with other commonly used lensing constraints~\citep{Young2022IFU3DLensReconstruction}. Low continuum flux contamination to narrow nebular lines is another practical advantage in crowded cluster fields.

\paragraph{5) Effects of Dark Matter Substructure:}
The local lens map is expected to be perturbed by DM subhalos, which produces “wiggles” in the critical curve~\citep{Dai2018Abell370, Williams2023FlashlightsDMSubhalo, Ji25SubhaloEffectOnCaustics, Perera2025arXiv251104748P}. While a simple polynomial deflection field with enforced smoothness may not capture such perturbation, significant misfit with high-quality JWST/NIRSpec IFU data in the future may be a hint for the subhalos. To detect the effect of subhalos, the local lens mapping may be extended to allow sinusoidal spatial variations, which can be fit to data.

\section{Conclusion}
\label{sec:concl}

We have developed a method to determine the local lens map in the vicinity of a lensing critical curve intersecting a giant arc. 
The method relies on mapping the line-of-sight velocity field across the arc using emission-line data from IFU spectroscopy. 
For disk galaxies, it combines a local polynomial deflection field with a source-plane disk rotation model. 
Assuming that the local deflection field varies smoothly, the method infers the critical-curve location together with uncertainty estimates. 
Through mock tests, we find that applying Wiener deconvolution to IFU datacubes before velocity measurement substantially mitigates the bias caused by instrumental PSF effects.

For the specific case of the Dragon Arc in the Abell 370 lensing field, we further validate the method using mock observations.
We construct a mock source galaxy with clumpy nebular emissions representative of a star-forming spiral galaxy, map it through the lens model to generate an image-plane configuration resembling the Dragon Arc, and then perform Bayesian inference on the mock data.

Applying the method to archival VLT/MUSE IFU data for the Dragon Arc, we locate the critical curve in one highly magnified region of the arc with a \(1\sigma\) uncertainty of \(0.23\arcsec\). 
Near the middle of this arc segment, our inferred band is closest to the WSLAP+ critical curve and has an orientation most similar to that of the BUFFALO model, while it disfavors the locations predicted by several other global lens models. 
The band is sufficiently narrow to distinguish our inferred critical-curve location from the predictions of several other global lens models. 
The inferred location is also qualitatively consistent with the expectation from microlensing by intracluster stars that most detected microlensing events of highly magnified stars should appear on the negative-parity side of the macro critical curve, although larger samples of lensed stars are needed for confirmation.

With its diffraction-limited PSF, JWST/NIRSpec IFU observations are expected to improve the constraint by a factor of \(\sim 2\)--\(3\). At high astrometric precision, a generalization of our approach that allows more degrees of freedom in the local lens map may uncover wiggly structures in the critical curve revealing small-scale dark matter substructures~\citep{Dai2018Abell370, Broadhurst2025FuzzyDM}.  Using wide-filter imaging data, \cite{Griffiths2021HamiltonObject} studied a ``folded'' lensed image of the ``Hamilton's Object'' and found locally smooth lens mapping on arcsecond scales near the critical curve, while smoothness on sub-arcsecond scales was unresolved. \cite{Perera2025arXiv251104748P} found tentative astrometric signals of small-scale perturbations for lensed compact surface brightness features on caustic crossing arcs. For uncovering such effects with kinematic mapping, the fidelity of the method will hinge on model flexibilities that can capture features in disk rotation and lensing perturbations from sub-galactic dark matter structures, for which future studies are needed. Finally, kinematic information can be included in a global lens modeling procedure that accounts for all available spatially resolved information on the extended giant arc~\citep{Eid2025Abell370LensModelFramework, Eid2025arXiv251111952E}.

\section*{acknowledgments}
The JWST data used in this work are obtained from the Mikulski Archive for Space Telescopes (MAST) at the Space Telescope Science Institute. The specific observations analyzed in this work can be accessed via \dataset[doi:10.17909/6cmj-mv58]{https://doi.org/10.17909/6cmj-mv58}. R.Z. is grateful for the Berkeley Physics International Education (BPIE) program during which this research project was initiated. The authors thank Masamune Oguri for helping with lens models. The authors would like to also thank Jessica Lu for useful discussions. L.D. acknowledges partial research grant support from the Alfred P. Sloan Foundation (Award Number FG-2021-16495), from the Frank and Karen Dabby STEM Fund in the Society of Hellman Fellows, and from the Office of Science, Office of High Energy Physics of the U.S. Department of Energy (Award Number DE-SC-0025293).
YF is supported by JSPS KAKENHI Grant Numbers JP22K21349 and JP23K13149.

\appendix
\section{Quantitative Comparison of Critical-Curve Locations}
\label{app:a}

In Section~\ref{sec:results}, we compare the critical-curve band inferred from kinematic mapping with predictions from independent global lens models. 
Here we quantify this comparison by measuring the slice-by-slice offsets between each global-model critical curve and our inferred \(1\sigma\) band. 
Across the ROI, we take horizontal slices and measure the RA position at which each global-model critical curve intersects each slice. 
For each slice, the offset is set to zero if the model critical curve lies within our inferred \(1\sigma\) band; otherwise, we assign the signed distance to the nearest edge of the band, with positive values corresponding to larger RA.

\begin{figure*}[t]
\centering
\includegraphics[width=\textwidth]{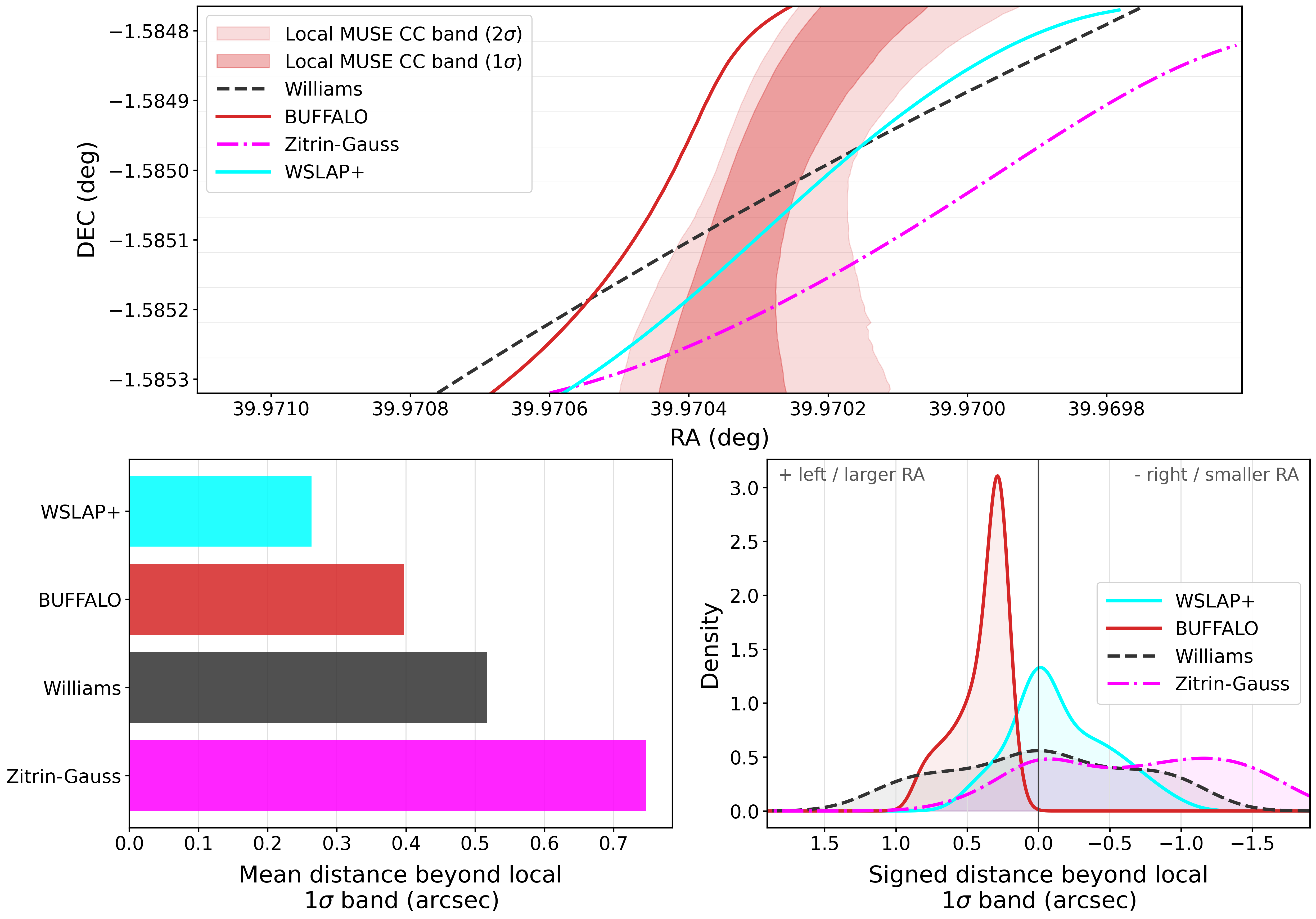}
\caption{
Quantitative comparison between the critical-curve band inferred from kinematic mapping and the critical curves predicted by independent global lens models. 
Top: inferred \(1\sigma\) and \(2\sigma\) critical-curve bands compared with the \texttt{WSLAP+}, \texttt{BUFFALO}, \texttt{Williams}, and \texttt{Zitrin-Gauss} models. 
This panel is shown with matched angular scales in RA and DEC. 
Bottom left: mean distance of each global-model critical curve outside the inferred \(1\sigma\) band. 
Bottom right: density distributions of the slice-by-slice signed distances outside the \(1\sigma\) band.
}
\label{fig:models_quantification}
\end{figure*}

With this metric, the \texttt{WSLAP+} critical curve is closest to our inferred band, with a mean distance outside the \(1\sigma\) band of \(0.263\arcsec\). 
The corresponding values are \(0.397\arcsec\) for \texttt{BUFFALO}, \(0.517\arcsec\) for \texttt{Williams}, and \(0.747\arcsec\) for \texttt{Zitrin-Gauss}. 
This provides a quantitative version of the visual comparison discussed in Section~\ref{sec:results}.

\section{Rotation Velocity Prior and Critical Curve Constraints}
\label{app:b}

To verify that our derived constraint on the critical curve is not compromised by the limited capability to determine the global disk rotation velocity $v_{\rm rot}$, we perform an analysis assuming a narrow prior $v_{\rm rot}=195$–$205\,$km/s. This test confirms that an accurate inference on $v_{\rm rot}$ is not necessary for correctly locating the critical curve.

\begin{figure}[h]
    \centering
    \includegraphics[width=\columnwidth]{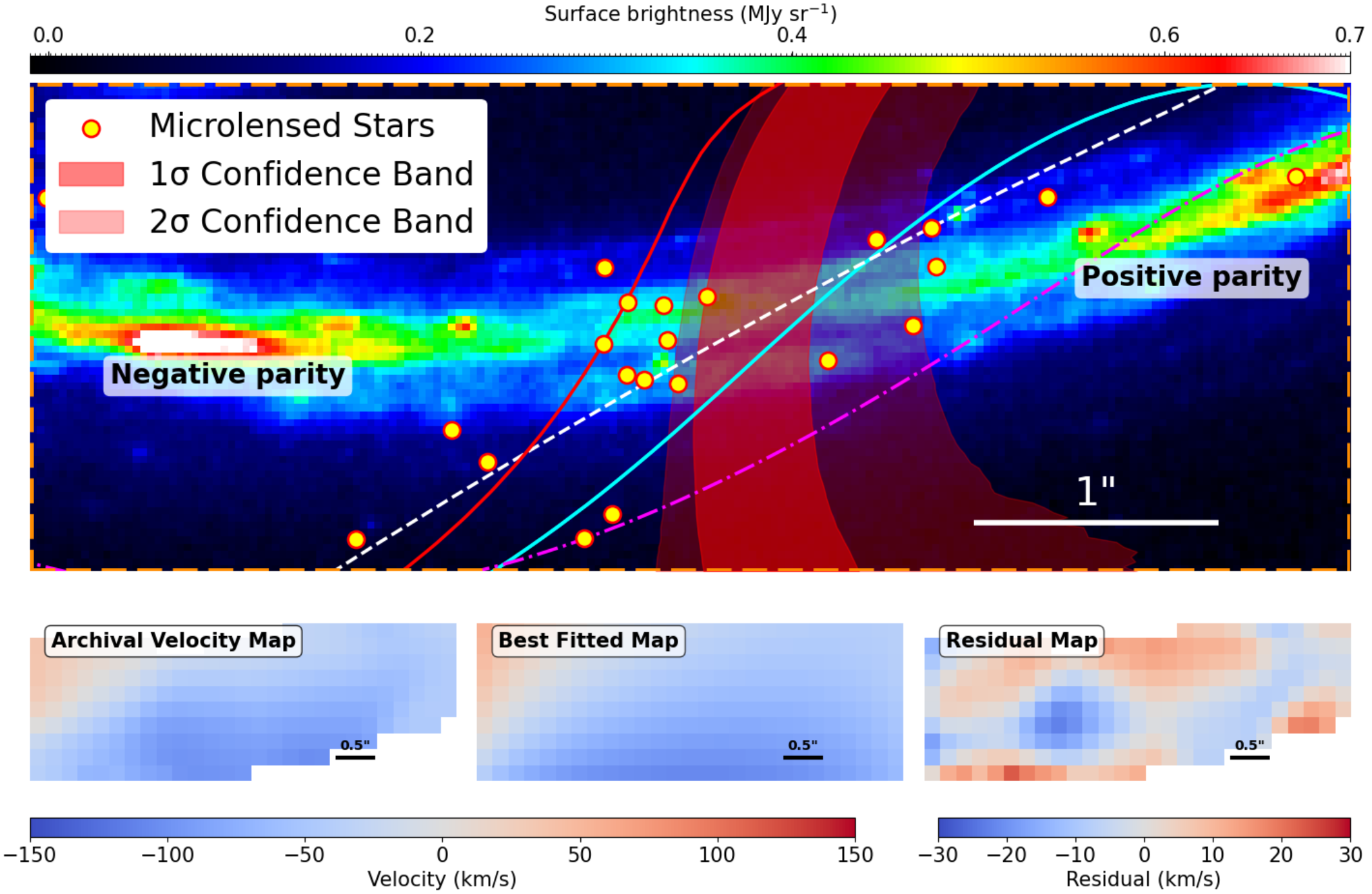}
    \caption{Critical curve constraints derived using a narrow prior on the rotation velocity ($v_{\rm rot} = 195$–$205\,$km/s). The resulting confidence band is nearly identical to that obtained with a broad prior (see Figure~\ref{fig:archival}), demonstrating the robustness of the critical curve localization. The mean half-width of the $1\sigma$ ($2\sigma$) confidence band is $0.23\arcsec$ ($0.52\arcsec$). Critical curves predicted by the same four global lens models as described in Figure~\ref{fig:Nircam} are also drawn: \texttt{Williams} v4.1 \citep[dashed white;][]{William_v4.1_model}, \texttt{Zitrin-Gauss} v1.0 \citep[dash-dotted purple;][]{Zitrin2009lensingltm, Zitrin2013lensingmodel}, \texttt{BUFFALO} v1.0 \citep[red solid;][]{Bufflomodel}, and the \texttt{WSLAP+} model \citep[cyan solid;][]{2025DiegoModel}.}
    \label{fig:appendix_vrot_prior}
\end{figure}

Figure~\ref{fig:appendix_vrot_prior} shows the resulting critical curve constraint. The confidence band derived with the narrow $v_{\rm rot}$ prior is very similar to that derived with the broad prior $v_{\rm rot}=150$–$350\,$km/s as shown in Figure~\ref{fig:archival}. The middle part of the band is unchanged, with only modest offsets in the upper and lower outskirts of the arc where kinematic information is inadequate.

The mean half width of the $1\sigma$ ($2\sigma$) confidence band is $0.23\arcsec$ ($0.52\arcsec$), virtually the same as the $0.23\arcsec$ ($0.48\arcsec$) width derived using the broad prior. We also note that the velocity residual maps look nearly identical between the two cases. 

The inferred parameters from both analyses are compared in Table~\ref{tab:app_parameter_comparison}. While the disk orientation parameters ($i$, $\phi$) show slight differences, the quality of the best-fit, as indicated by the reduced $\chi^2$ value, appears nearly identical. Most importantly, the constraint on the critical curve is consistent between the two cases. This suggests that our ability to constrain the critical curve is robust, even though, perhaps unsurprisingly, fitting a small part of the disk near the lensing caustic is inadequate for correctly reconstructing the global kinematic profile of the galaxy.

\begin{table}[h]
\centering
\caption{Comparison of derived disk parameters and critical curve constraints using broad versus narrow priors for the rotation velocity $v_{\rm rot}$. The position angle $\phi$ is measured from North ($0^\circ$) to East ($+90^\circ$).}
\label{tab:app_parameter_comparison}
\begin{tabular}{c|c|c}
\toprule
 & \textbf{Broad Prior} & \textbf{Narrow Prior} \\
 & (150--350 km/s) & (195--205 km/s) \\
\midrule
\textbf{Parameters (68\% C.I.)} & & \\
$i$ [$^\circ$] & $(57,\,78)$ & $(69,\,80)$ \\
$\phi$ [$^\circ$] & $(-43,\,-32)$ & $(-46,\,-37)$ \\
$v_{\rm rot}$ [km~s$^{-1}$] & $(267,\,329)$ & $(198,\,204)$ \\
\cmidrule{1-3}
\textbf{Model Performance} & & \\
$\chi^2$/dof (best fit) & $0.93$ & $0.95$ \\
$1\sigma$ band [arcsec] & $0.23$ & $0.23$ \\
$2\sigma$ band [arcsec] & $0.48$ & $0.52$ \\
\bottomrule
\end{tabular}
\end{table}

\bibliographystyle{aasjournal}
\bibliography{refs}
\end{document}